\documentclass[a4paper, 12pt, openbib ]{article}
\usepackage{times}
\usepackage{pifont}
\usepackage{amsmath}
\usepackage{amssymb}
\usepackage{amsfonts}
\usepackage{graphicx}
\usepackage{floatflt}
\usepackage{geometry}
\usepackage{layout}
\usepackage{enumerate}
\usepackage{bm}
\usepackage{tikz}
\usetikzlibrary{patterns}
\def\ner{\boldsymbol}
\usepackage{amsthm}
\usepackage{thmtools}
\def\tfract#1/#2{{\textstyle{\raise0.8pt\hbox{$\scriptstyle#1$}\over%
\hbox{\lower0.8pt\hbox{$\scriptstyle#2$}}}}}
\def\mezzo{\tfract 1/2 }
\def\imezzo{\tfract -i/2 }

\def\dueterzi{\tfract 2/3}

\def\sesto{\tfract 1/6 }

\def\radi2k{\tfract 1/{\sqrt {2k}} }
\def\der{\partial }
\def\cvd{\vbox{\hrule \hbox to 9 pt {\vrule height 9 pt \hfil \vrule} \hrule}}
\def\downnormalfill{$\,\,\vrule depth4pt width0.4pt
\leaders\vrule depth 0pt height0.4pt\hfill\vrule depth4pt width0.4pt\,\,$}
\def\WT#1{\mathop{\vbox{\ialign{##\crcr\noalign{\kern3pt}
      \downnormalfill\crcr\noalign{\kern0.8pt\nointerlineskip}
      $\hfil\displaystyle{#1}\hfil$\crcr}}}\limits}

\def\be{\begin{equation}}
\def\ee{\end{equation}}
\def\bes{\begin{equation*}}
\def\ees{\end{equation*}}
\def\bea{\begin{eqnarray}}
\def\eea{\end{eqnarray}}
\def\beas{\begin{eqnarray*}}
\def\eeas{\end{eqnarray*}}
\def\ba{\begin{array}{rcl}}
\def\ea{\end{array}}
\def\der{\partial}
\numberwithin{equation}{section}
\def\go{\leavevmode \raise.3ex\hbox{$\scriptscriptstyle \langle\!\langle\!  $}%
~\ignorespaces}
\def\gf{\relax \ifhmode \unskip~\else \leavevmode \fi \raise.3ex\hbox{$\! \scriptscriptstyle\rangle\!\rangle\, $}}
\title{
{\Large  \bf Perturbative BF theory}
{\vskip 0.6 truecm}} 

\author{{\large Enore Guadagnini and Federico Rottoli} \\  {\normalsize {~}} \\  {\normalsize  Dipartimento di Fisica {\it E. Fermi} dell'Universit\`a di Pisa,} \\  {\normalsize and INFN Sezione di Pisa,} \\ {\normalsize   Largo B. Pontecorvo  3, 56127 Pisa, Italy.} 
}

\date{}

\begin{document}

\maketitle 

\vskip 0.7 truecm 

\begin{abstract}

We consider a superrenomalizable gauge theory of topological type, in which the  structure group is equal to the inhomogeneous   group $ISU(2)$. The generating functional of the correlation functions of the gauge fields is derived and its connection with the generating functional of the Chern-Simons theory is discussed.   The complete renomalization  of this model defined in $\mathbb{R}^3$ is presented. The structure of the $ISU(2)$ conjugacy classes is determined. Gauge invariant observables are defined by means of appropriately normalized traces of $ISU(2)$ holonomies associated with oriented, framed and coloured knots.    The perturbative  evaluation of the Wilson lines expectation values is investigated and the up-to-third-order contributions  to the perturbative expansion of the  observables,  which correspond to knot   invariants,  are produced.  The general dependence of the knot observables on the framing is worked out. 

 \end{abstract}

\vskip 1.2 truecm

\section{Introduction}

Among the quantum field theory models of topological type \cite{BB}, the so-called BF theory \cite{HO,KR,MP,BT} has been proposed in order to describe several different phenomena. 

The gauge structure group of the BF theory naturally suggests  possible connections with (2+1) gravity \cite{HC,PLE,MDM,EW,GIU,GU1,HU,KRA,DUR,SPE,INO,JIN,DMP,ISW}, and applications of the BF formalism in the context of loop quantum gravity have also been studied   \cite{EZA,BAE,LIV,CIA,MER,DL,BMR}. Generalizations of the BF models  in higher dimensions  have  been considered \cite{GMS,MOM,PU,NOB,CAR,HAN,MAU,ALI,WIN,PFE,VAL,GO,ESC,ITS,SAN,GCM}. 
Quite recently, the use of the BF field theory has been envisaged for the description of topological effects in condensed matter \cite{YCM,MLS,ANN,VS,ABC,TCN,CDT,ZJW,YOU}. 

Various BF quantization procedures have been examined  \cite{IKE,CGR,BRA,IVA,SMA,FER,GHA,TON,BOR,BIZ}  and the renormalizability of the theory has been proved by power counting and general arguments \cite{MS, DBR, LUC,FUC,OSW,CLI}.  The BF model is actually a superrenormalizable theory; nevertheless, the explicit  renormalization ---with specified normalization conditions--- has never been produced.  One of the purposes of the present article is precisely to provide the complete renormalization of the nonabelian BF theory in ${\mathbb R}^3$. 

The definition and computation of topological invariants  \cite{J,BER,CAT,CAI,LAU,TAT,NOR,GAD,PAV,ERN} are central issues in the  BF model. The observables that we propose have not been considered in literature.  We shall demonstrate that  the appropriately normalized  traces of the expectation values of the holonomies ---for the inhomogeneous group $ISU(2)$---   associated with oriented framed knots in ${\mathbb R}^3$   are well defined.  The first three orders of the  perturbative computation of these observables are presented. 

Let us recall that the solution of the abelian BF theory in generic closed oriented 3-manifolds has been produced by Mathieu and Thuillier \cite{MFR,FMU,PMS}.  
In the present paper we shall concentrate on the perturbative approach to the nonabelian BF theory in ${\mathbb R}^3$ with structure group $ISU(2)$. 

The Lie algebra of the inhomogeneous group $ISU(2)$ can be interpreted as a particular extension of the  $SU(2)$ algebra which, in the quantum mechanics description of one particle moving in ${\mathbb R}^3$, is obtained by the introduction of the  three components $P^a$ of the momentum in addition to the three components $J^a$ of the angular momentum. 
The corresponding $ISU(2)$ connection has then six components ${\cal A}_\mu = A^a_\mu (x) J^a + B^a_\mu (x) P^a $. The most general action in ${\mathbb R}^3$ which is  $ISU(2)$ gauge invariant and metric-independent contains two different terms: the first term $\int B^a \wedge F^a(A)$ ---where $F^a(A)$ are the angular momentum components of the curvature--- gives the name to the model  and the second term $\int {\rm Tr}( A\wedge d A  + i \dueterzi A\wedge A \wedge A)$ coincides with the  Chern-Simons action  for the $SU(2)$ subgroup. 

 Section~2 contains the fundamentals of the perturbative approach for the computation of the BF correlation functions of the connection in the Landau gauge. The general  structure of the connected Feynman diagrams is worked out. The computation of the generating functional of the connected correlation functions to all orders of perturbation theory is presented and its Chern-Simons relationship is discussed in Section~3. The complete renormalization of the BF theory is given in Section~4. It is shown that the theory is superrenormalizable, and only six one-loop diagrams   need to be examined.   These one-particle-irreducible diagrams  concern the two-point function and the three-point proper vertex of the connection. It is shown that, as in the case of the Chern-Simons theory, the two-point function of the connection does not receive loop corrections  and therefore the bare propagator coincides with the dressed propagator. 
 
 In order to introduce Wilson line observables in the BF model,  certain unitary representations of $ISU(2)$ are described in Section~5. Since the group   $ISU(2)$ is noncompact, these nontrivial representations are infinite dimensional. Wilson line operators are defined by means of  normalized traces of the $ISU(2)$ holonomies associated with oriented knots.   For completeness, the  classical traces of the $ISU(2)$ conjugacy classes are described in Section~6. The proof that the BF expectation values of the Wilson line operators are well defined is contained in Section~7. It is shown that, since the correlation functions of the connection are invariant under global $ISU(2)$ transformations, the expectation value of a knot holonomy is a function of the Casimir operators of $ISU(2)$.    This implies that the BF mean values of the Wilson line operators are well defined and describe topological invariants for oriented and framed  knots in ${\mathbb R}^3$. 
 
 The perturbative computation of the knot observables   up to the third order in powers of $\hbar$ is described in Section~8.   The knot invariants that are found at first and second order correspond to the knot invariants that also appear in the Chern-Simons theory. While, at the third order of perturbation theory,   the BF and Chern-Simons knot invariants differ.   A proof  that the entire framing dependence of the knot observables is completely determined  by an overall multiplicative factor is given. This factor is the exponential of the linking number between the knot and its framing multiplied by the combination of the quadratic Casimir operators which is determined by the two point function of the connection.  
Section~9 contains the conclusions.

\section{Fields, lagrangian and diagrams}

The fundamental fields of the so-called BF theory \cite{BB,HO,KR,MP,BT,EW} are given by  the components of the $ISU(2)$ connection 
\be
{\cal A} = {\cal A}_\mu(x) dx^\mu = 
\left \{  A^a_\mu (x) J^a + B^a_\mu (x) P^a  \right \} dx^\mu \; , 
\label{2.1}
\ee
where the generators $(J^a , P^a ) $ (with $a=1,2,3$) of the algebra of $ISU(2)$ satisfy the commutation relations 
\be
[J^a , J^b] = i \epsilon^{abc} J^c \quad , \quad [J^a , P^b] = i \epsilon^{abc} P^c \quad , \quad  
 [P^a , P^b] = 0 \; . 
 \label{2.2}
\ee
Let us consider the BF model defined in $\mathbb{R}^3$.  Gauge transformations act as 
\be
{\cal A} \longrightarrow {\cal A}^\Omega = \Omega^{-1} {\cal A} \Omega -i \Omega^{-1} \, d\Omega \; , 
\label{2.3} 
\ee
where $\Omega : {\mathbb R}^3 \rightarrow ISU(2)$. When $\Omega \simeq 1 + i \beta^a J^a + i \eta^a P^a $, the infinitesimal gauge transformations take the form 
\bea
A_\mu^a \rightarrow A_\mu^a + \Delta A_\mu^a  \quad &,& \quad  \Delta A_\mu^a = \der_\mu \beta^a - \epsilon^{abc} A_\mu^b \beta^c \nonumber \\
B_\mu^a \rightarrow B_\mu^a + \Delta B_\mu^a \quad &,& \quad \Delta B^a_\mu = \der_\mu \eta^a - \epsilon^{abc} A_\mu^b \eta^c - \epsilon^{abc} B_\mu^b \beta^c \; . 
\label{2.4}
\eea 
The components of the curvature are given by 
\bea
{\cal F}_{\mu \nu} &=& - i [\der_\mu + i {\cal A}_\mu , \der_\nu + i {\cal A}_\nu ]\nonumber \\ 
&=& F^a_{\mu \nu }(A) J^a + \left ( D_\mu (A) B_\nu - D_\nu (A)B_\mu \right )^a P^a \; , 
\label{2.5}
\eea
in which 
\be
F^a_{\mu \nu }(A) = \der_\mu A_\nu^a - \der_\nu A_\mu^a  - \epsilon^{abc} A^b_\mu  A^c_\nu \; , 
\label{2.6}
\ee
and 
\be
\left ( D_\mu (A) B_\nu \right )^a = \der_\mu B_\nu^a - \epsilon^{abc} A^b_\mu B^c_\nu \; . 
\label{2.7}
\ee
The action  of the BF theory in ${\mathbb R}^3$  is the sum of the two  metric-independent terms which are separately invariant under $ISU(2)$ transformations (\ref{2.4}) 
\be
S = \int d^3 x \, \epsilon^{\mu \nu \lambda}  \left \{ \mezzo  \, B^a_\mu F^a_{\nu \lambda} (A) + g  \left [ \mezzo A^a_\mu \der_\nu A^a_\lambda
 - \sesto \epsilon^{abc }\, A^a_\mu A^b_\nu A^c_\lambda \right ] \right \} \; . 
 \label{2.8}
 \ee
 Without loss of generality, the overall normalization of the first term in expression (\ref{2.8}) can be taken to be $(1/2)$, because the $ISU(2)$ generators $P^a$ can be rescaled without any modification of the Lie algebra commutation relations (and consequently $B^a_\mu$ also can be rescaled).   The real parameter  $g$ is a dimensionless coupling constant  which multiplies the Chern-Simons lagrangian term 
 \be
 S_{CS} [A] =  \int d^3 x \, \epsilon^{\mu \nu \lambda} \left [ \mezzo A^a_\mu \der_\nu A^a_\lambda
 - \sesto \epsilon^{abc }\, A^a_\mu A^b_\nu A^c_\lambda \right ] \; . 
 \label{2.9}
 \ee 
 When $g =(k / 4\pi)$ with integer $k$, one also recovers invariance  under large gauge transformations, which anyway play no role in the perturbative approach to the theory. Note that, in order to discuss the renormalization of any gauge theory  model, all the possible lagrangian terms which are gauge invariant must be taken into account. This is why   the renormalization of the BF model requires that  both lagrangian terms ---shown in expression (\ref{2.8})--- must be included in the action. 

\subsection{Gauge fixing}
The gauge fixing procedure is implemented according to the BRST method \cite{BRS,TYU}. The BRST transformations \cite{WA} are given by 
\bea
\delta A_\mu^a =\der_\mu c^a - \epsilon^{abc} A_\mu^b c^c  \; &,& \; \delta B^a_\mu = \der_\mu \xi^a - \epsilon^{abc} A_\mu^b \xi^c
- \epsilon^{abc} B_\mu^b c^c \; , \nonumber \\
\delta c^a = \mezzo \epsilon^{abc} c^bc^c \quad  
, \quad \delta {\overline c}^a = M^a \; &,& \; 
\delta \xi^a = \epsilon^{abc} \xi^b c^c   \quad , \quad \delta {\overline \xi}^a = N^a \; ,  \\   \delta M^a =0 \; &,& \; \delta N^a =0 \; , \nonumber 
\label{2.10}  
\eea
where $\{  \xi^a , \overline \xi^a, c^a , {\overline c}^a   \}$ is the set of anticommuting ghosts and antighosts fields,  whereas $M^a , N^a $ represent the commuting auxiliary  fields.  In the Landau gauge,  the gauge-fixing and ghosts action terms are given by 
\bea
S_{\phi \pi } &=& \int d^3x \Bigl \{ M^a \der^\mu A^a_\mu + N^a \der^\mu B^a_\mu + \der^\mu  {\overline c}^a  (\der_\mu c^a - \epsilon^{abc} A_\mu^b c^c ) \nonumber \\ 
&& {\hskip 1.4 cm} + \der^\mu  {\overline \xi}^a  (\der_\mu \xi^a - \epsilon^{abc} A_\mu^b \xi^c
- \epsilon^{abc} B_\mu^b c^c) \Bigr \} \; ,  
\label{2.11} 
\eea
where the flat euclidean metric $g_{\mu \nu} = \delta_{\mu \nu}$  of ${\mathbb R}^3$ has been introduced in order to contract the vector indices.   The total action $ S_{TOT} = S + S_{\phi \pi }$ is invariant under BRST transformations. 

In order to recognize the structure constants of the $ISU(2)$ Lie algebra  in the gauge-fixing procedure, it is convenient to introduce the ghost field ${\cal C} = c^a J^a + \xi^a P^a$, the antighost field ${\overline {\cal C}} = {\overline \xi}^a J^a + {\overline c}^a P^a$ and the auxiliary field ${\cal N} = N^a J^a + M^a P^a$. The BRST transformations  take the form 
$$
\delta {\cal A}_\mu(x) = [D_\mu ({\cal A}), {\cal C} ] \quad , \quad \delta {\cal C} = \imezzo \{ {\cal C} , {\cal C} \} \quad , \quad  \delta {\overline {\cal C}} = {\cal N} \quad , \quad \delta {\cal N} = 0 \; , 
$$
and $S_{\phi \pi }$ can be written as 
$$
S_{\phi \pi } = \delta \int d^3x \, \langle {\overline {\cal C}}\, \der^\mu \! {\cal A}_\mu \rangle^{JP} \; , 
$$
where the bracket $\langle \cdot \cdot \rangle^{JP}$ denotes the non-degenerate bilinear form \cite{EW} on the $ISU(2)$ algebra 
$$
\langle J^a P^b \rangle^{JP} = \delta^{ab} \quad , \quad \langle J^a J^b \rangle^{JP} = 0 = \langle P^a P^b \rangle^{JP} \; . 
$$

\subsection{Propagators}

The Green functions of the differential operators acting on the fields ---and entering the quadratic parts of $S_{TOT}$ in powers of the fields--- determine the form of the fields propagators. As far as the bosonic fields are concerned, the nonvanishing components of the propagators are given by 
\bea
\WT{A^a_\mu (x) B}\! \null^b_\nu(y) &=& \delta^{ab} \int {d^3k \over (2 \pi )^3} \, e^{i k(x-y)} \, \epsilon_{\mu \nu \lambda} {k^\lambda \over k^2} = - i  \delta^{ab} \epsilon_{\mu \nu \lambda} \der^\lambda \Delta (x-y)  \; , \nonumber \\ 
\WT{B^a_\mu (x) B}\! \null^b_\nu(y) &=& - g \, \delta^{ab} \int {d^3k \over (2 \pi )^3} \, e^{i k(x-y)} \, \epsilon_{\mu \nu \lambda} {k^\lambda \over k^2} =  i g \delta^{ab} \epsilon_{\mu \nu \lambda} \der^\lambda \Delta (x-y) \; , 
\label{2.12}
\eea
and
\bea
\WT{A^a_\mu (x)M}\! \null^b(y) &=& \delta^{ab} \int {d^3k \over (2 \pi )^3} \, e^{i k(x-y)} \, {k_\mu \over k^2} = -i  \delta^{ab} \der_\mu  \Delta (x-y) 
\; , \nonumber \\ 
\WT{B^a_\mu (x)N}\! \null^b(y) &=&  \delta^{ab} \int {d^3k \over (2 \pi )^3} \, e^{i k(x-y)} \, {k_\mu \over k^2} = -i  \delta^{ab} \der_\mu  \Delta (x-y)  \; . 
\label{2.13}
\eea
For the anticommuting fields one gets 
\be
\WT{c^a (x) \, {\overline c}}\! \null^b(y) = 
\WT{\xi^a (x) \, {\overline \xi}}\! \null^b(y) = 
i \delta^{ab} \int {d^3k \over (2 \pi )^3} \, e^{i k(x-y)} \, {1\over k^2} = i \delta^{ab} \Delta(x-y)\; . 
\label{2.14}
\ee

\subsection{Structure of the  diagrams}
The Feynman diagrams of the BF theory, which is  defined by the action $S_{TOT} = S + S_{\phi \pi }$ in ${\mathbb R}^3$, have quite peculiar properties that we shall now discuss.  Let us consider the generating functional $W[J,K]$ of the connected correlation functions of the components of the connection 
\be
e^{i W[J,K]} = \langle   e^{i \int d^3x (J^a_\mu A^a_\mu + K^a_\mu  B^a_\mu)}  \rangle ={\int D\hbox{(fields) } e^{iS_{TOT}} \, e^{i \int d^3x (J^a_\mu A^a_\mu + K^a_\mu  B^a_\mu)}\over 
\int D\hbox{(fields) } e^{iS_{TOT}}  } \; , 
\label{2.15}
\ee
where $J^a_\mu(x)$ and $K^a_\mu (x)$ are classical sources. We are interested in $ W[J,K]$ because in the following sections we shall consider the expectation values of observables which are functions of the fields $A^a_\mu$ and $B^a_\mu$ exclusively.  In this section we consider the combinatorial structure ---which is determined by the Wick contractions--- of the Feynman diagrams contributing to $W[J,K]$. The renormalization will be treated in Section~4.  
The first issue to be discussed is an extension of the  Oda and Yahikozawa observation presented  in \cite{OY}. 

\newtheorem{prop}{Proposition}
\begin{prop}
The entire generating functional $W[J,K]$ is given by the sum of connected Feynman diagrams with no loops   and with one loop only,  
\be
W[J,K]= W_{0} [J,K] + W_{1}[J,K] \; . 
\label{2.16}
\ee
The contribution $W_0 [J,K]$ of the tree-level Feynman diagrams can be decomposed into the sum of two terms, 
\be
 W_{0} [J,K] =  U[K] + \int d^3x J^a_\mu (x) H_\mu ^a [K](x)\; , 
 \label {2.17}
\ee 
in which $U[K]$ and $H_\mu ^a [K](x)$ only depend on $K^a_\mu$.  The term  $U[K]$  is linear in $g$ whereas  $H_\mu ^a [K](x)$  does  not depend on $g$. The contribution  $W_{1}[J,K]$  of the  one-loop diagrams does not depend on $g$ and does not depend on $J^a_\mu$, 
\be
W_{1}[J,K] = W_1[K] \; . 
\label{2.18}
\ee 
\end{prop}

\smallskip

\noindent {\bf Proof.}  Let us first consider the connected tree level diagrams which contribute to $W[J,K]$. Diagrams which do not contain interaction vertices obviously satisfy equation  (\ref{2.17}) because  $\WT{AA} =0 $ and the nonvanishing components of the propagators are shown in equation (\ref{2.12}). So let us now concentrate on diagrams  which contain interaction vertices, which are  of the type $BAA$ or of the type $AAA$; it is convenient to recover all these diagrams in three steps. 

\begin{enumerate}
\item The diagrams constructed with  $BAA$ interaction vertices  and $\WT{AB}$  propagators exclusively are called the basic diagrams; these are the diagrams that remain in the $g \rightarrow 0 $ limit. They contain one power of the field  $J^a_\mu$ and may contain an arbitrary  number bigger than unit  of $K^a_\mu $ fields. One example is shown in Figure~1(a). Indeed, each  tree diagram is  obtained by combining subdiagrams called ``branches". One branch is a one-dimensional ordered sequence of vertices connected by propagators, as  shown in Figure~1(b).  
 Note that the external  lines of Figure~1(b) correspond to field components and do not represent propagators; in particular,  one branch diagram necessarily  has  external legs corresponding to one $B^a_\mu$ field and several $A^a_\mu$ fields.

 \vskip 0.4 truecm

\centerline {
\begin{tikzpicture} [scale= 0.8] 
%
\draw[very thick] (1,2) -- (3,2);
\draw[very thick] (0,1) -- (1,2);
\draw[very thick] (0,3) -- (1,2);
\draw[very thick] (3,2) -- (4,1);
\draw[very thick] (3,2) -- (4,3);
%
\node at (1.3,2.24) {\footnotesize $  B $};
\node at (0.79,1.52) {\footnotesize $  A $};
\node at (0.82,2.58) {\footnotesize $  A $};
\node at (2.7,2.24) {\footnotesize $  A $};
\node at (3.1,1.52) {\footnotesize $  A $};
\node at (3.2,2.58) {\footnotesize $  B $};
\node at (0,3) {$\bullet$};
\node at (0,1) {$\bullet$};
\node at (4,1) {$\bullet$};
\node at (4,3) {$\bullet$};
\node at (-.4,1) {$K$};
\node at (-.4,3.1) {$K$};
\node at (4.4,1) {$K$};
\node at (4.4,3.1) {$J$};
\node at (2,0) {$(a)$};
\draw[very thick] (14,1) -- (14,3);
\draw[very thick] (10,2) -- (10,3);
\draw[very thick] (8,1) -- (8,3);
\draw[very thick] (8,2) -- (10.5,2);
\draw[very thick] (11.5,2) -- (14,2);
\draw[very thick] (12,2) -- (12,3);
\node at (11.04,2) {\Large $\cdots$}; 
\node at (7.7,3) {\footnotesize $  A $};
\node at (7.7,1) {\footnotesize $  A $};
\node at (8.3,2.24) {\footnotesize $  B $};
\node at (9.7,2.24) {\footnotesize $  A $};
\node at (10.3,2.24) {\footnotesize $  B $};
\node at (11.7,2.24) {\footnotesize $  A $};
\node at (12.3,2.24) {\footnotesize $  B $};
\node at (13.7,2.24) {\footnotesize $  A $};
\node at (9.7,3) {\footnotesize $  A $};
\node at (11.7,3) {\footnotesize $  A $};
\node at (13.7,3) {\footnotesize $  B $};
\node at (13.7,1) {\footnotesize $  A $};
\node at (11,0) {$(b)$};
\end{tikzpicture}
}

\vskip 0.3 truecm
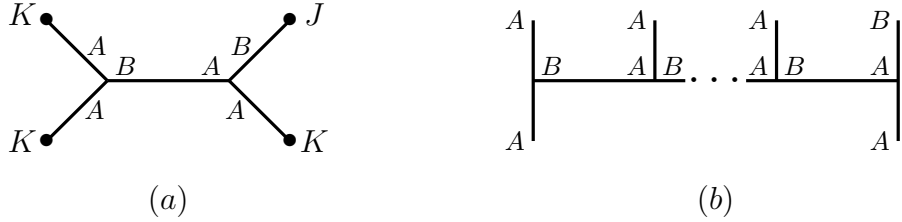
\centerline {{Figure~1.} {$(a)$ Example of basic diagram. $(b)$ Branch diagram.}}

\vskip 0.3 truecm

\item By adding the possibility of using also $\WT{BB}$ propagators, the new diagrams only contain one extra $\WT{BB}$ propagator  ---with respect to the basic diagrams of the previous case--- and then they are of first order in powers of $g$ and do not depend on $J^a_\mu$. See for instance Figure~2(a). The  $\WT{BB}$ propagator may correspond to one internal line in the Feynman diagrams, or to an external leg of the diagrams. 

 \vskip 0.3 truecm

\centerline {
\begin{tikzpicture} [scale= 0.8] 
%
\draw[very thick] (1,2) -- (3,2);
\draw[very thick] (0,1) -- (1,2);
\draw[very thick] (0,3) -- (1,2);
\draw[very thick] (3,2) -- (4,1);
\draw[very thick] (3,2) -- (4,3);
\node at (1.3,2.24) {\footnotesize $  B $};
\node at (0.79,1.52) {\footnotesize $  A $};
\node at (0.82,2.58) {\footnotesize $  A $};
\node at (2.7,2.24) {\footnotesize $  B $};
\node at (3.1,1.52) {\footnotesize $  A $};
\node at (3.15,2.58) {\footnotesize $  A $};
\node at (0,3) {$\bullet$};
\node at (0,1) {$\bullet$};
\node at (4,1) {$\bullet$};
\node at (4,3) {$\bullet$};
\node at (-.4,1) {$K$};
\node at (-.4,3.1) {$K$};
\node at (4.4,1) {$K$};
\node at (4.4,3.1) {$K$};
\node at (2,0) {$(a)$};
\begin{scope}[shift={(9,0)}]
\draw[very thick] (1,2) -- (3,2);
\draw[very thick] (0,1) -- (1,2);
\draw[very thick] (0,3) -- (1,2);
\draw[very thick] (3,2) -- (4,1);
\draw[very thick] (3,2) -- (4,3);
\node at (1.3,2.24) {\footnotesize $  B $};
\node at (0.79,1.52) {\footnotesize $  A $};
\node at (0.82,2.58) {\footnotesize $  A $};
\node at (2.7,2.24) {\footnotesize $  A $};
\node at (3.1,1.52) {\footnotesize $  A $};
\node at (3.15,2.58) {\footnotesize $  A $};
\node at (0,3) {$\bullet$};
\node at (0,1) {$\bullet$};
\node at (4,1) {$\bullet$};
\node at (4,3) {$\bullet$};
\node at (-.4,1) {$K$};
\node at (-.4,3.1) {$K$};
\node at (4.4,1) {$K$};
\node at (4.4,3.1) {$K$};
\end{scope}
\node at (11,0) {$(b)$};
\end{tikzpicture}
}

\vskip 0.2 truecm
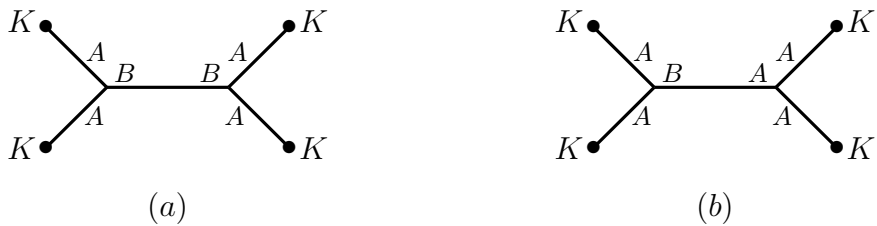
\centerline {{Figure~2.} {$(a)$ Example of  diagram with one $\WT{BB}$ propagator. $(b)$ Diagram with one $AAA$ vertex.}}

\vskip 0.4 truecm

\item Finally, by incorporating  the additional possibility of including also vertices of the  $AAA$ type, the new diagrams only contain one extra $AAA$ vertex with respect to the basic diagrams, they are linear in $g$  and do not depend on $J^a_\mu$, as shown in the example of Figure~2(b). 
\end{enumerate}

\noindent As a result, the set of all the connected tree-level diagrams contains diagrams which are linear in $g$ and do not depend on $J^a_\mu$ and diagrams which  linearly depend on  $J^a_\mu$ and do not depend on $g$. This concludes the proof of equation (\ref{2.17}). 

Let us now consider the one-loop connected diagrams entering $W[J,K]$. As shown in the example of Figure~3(a), connected diagrams with one loop of ghosts only depend on the source field $K^a_\mu$ because of the particular structure of the ghosts couplings (\ref{2.11}).  As far as the diagrams without a ghost loop are concerned, by cutting one internal propagator of each one-loop diagram one can open the loop and obtain  a connected zero-loop  diagram. In view of the result (\ref{2.17}), the broken propagator was necessary of the $AB$ type. Consequently, also each one-loop diagram with no ghost propagators  does not  depend on the $J^a_\mu$ field and does not depend on $g$, see the example of Figure~3(b). This concludes the proof of equation (\ref{2.18}).  

 \vskip 0.3 truecm

\centerline {
\begin{tikzpicture} [scale= 0.8] 
%
\draw[very thick, dashed ] (1.2,1.2) -- (2.8,1.2);
\draw[very thick, dashed ] (1.2,1.2) -- (2,2.7);
\draw[very thick, dashed ] (2.8,1.2) -- (2,2.7);
\draw[very thick] (0.5,0.5) -- (1.2,1.2);
\draw[very thick] (3.5,0.5) -- (2.8,1.2);
\draw[very thick] (2,3.5) -- (2,2.7);
\node at (0.5,0.5) {$\bullet$};
\node at (3.5,0.5) {$\bullet$};
\node at (2,3.5) {$\bullet$};
\node at (0,0.5) {$K$};
\node at (4,0.5) {$K$};
\node at (2.5,3.6) {$K$};
\node at (2,-0.6) {$(a)$};
\begin{scope}[shift={(8,0)}]
\draw[very thick ] (1.2,1.2) -- (2.8,1.2);
\draw[very thick ] (1.2,1.2) -- (2,2.7);
\draw[very thick ] (2.8,1.2) -- (2,2.7);
\draw[very thick] (0.5,0.5) -- (1.2,1.2);
\draw[very thick] (3.5,0.5) -- (2.8,1.2);
\draw[very thick] (2,3.5) -- (2,2.7);
\node at (1.07,1.56) {\footnotesize $  B $};
\node at (1.52,2.55) {\footnotesize $  A $};
\node at (2.46,2.55) {\footnotesize $  B $};
\node at (2.96,1.56) {\footnotesize $  A $};
\node at (1.39,0.96) {\footnotesize $  A $};
\node at (2.61,0.96) {\footnotesize $  B $};
\node at (0.5,0.5) {$\bullet$};
\node at (3.5,0.5) {$\bullet$};
\node at (2,3.5) {$\bullet$};
\node at (0,0.5) {$K$};
\node at (4,0.5) {$K$};
\node at (2.5,3.6) {$K$};
\end{scope}
\node at (10,-0.6) {$(b)$};
\end{tikzpicture}
}

\vskip 0.3 truecm
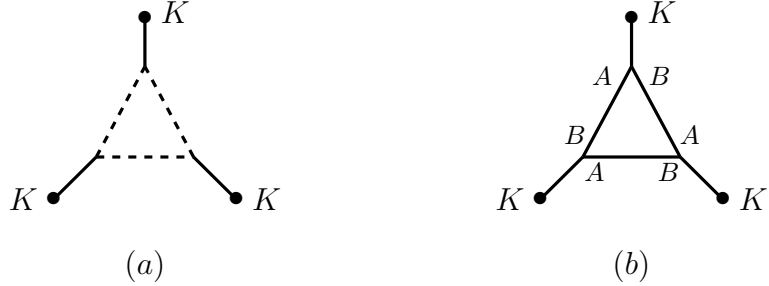
\centerline {{Figure~3.} {$(a)$ Diagram with one ghost loop. $(b)$ One-loop diagram without ghost propagators.}}

\vskip 0.4 truecm

\noindent Finally, there are no connected diagrams with two or more loops contributing to $W[J,K]$ because all the one-loop diagrams have external legs corresponding to the $A$ field and  the component $\WT{AA}$ of the propagator  is vanishing. \hfill \cvd

\medskip

As a final remark, consider the contributions to $W[J,K]$ of the diagrams containing ghost loops. Since  only one-loop diagrams enter $W[J,K]$, all the corresponding possible subdiagrams  that have external ghost fields are tree-level diagrams (which are well defined and finite). Consequently, in discussing the renormalization of $W[J,K]$, the diagrams with external ghost fields  can be ignored. 

Let $X[A,B]$ be a function of the field  components $ A_\mu^a$ and $ B_\mu^a$. In the perturbative computation of the expectation value $\langle X[A,B] \rangle$, 
\be
\langle X[A,B] \rangle = {{\int D\hbox{(fields)} \, e^{iS_{TOT}} \, X[A,B]}\over 
{\int D\hbox{(fields) } \, e^{iS_{TOT}}  } } \; , 
\label{2.19}
\ee
the ghosts contributions are described by  diagrams with ghost loops. As shown in equation (\ref{2.14}), the nonvanishing components of the ghosts propagator  are of the  type $\WT{ c \; {\overline c}} $ or $\WT{ \xi \; {\overline \xi}  } $; therefore the lagrangian term $ \epsilon^{abd}\der^\mu {\overline \xi^a }(x) B^b_\mu (x)c^d(x)$ ---contained in $S_{TOT} = S + S_{\phi \pi}$--- does not  contribute to $\langle X[A,B] \rangle$.  

\section{Generating functionals and Chern-Simons relationship} 

In order to complete the description of the BF diagrams, in this section we derive the BF generating functional of the connected correlation functions and discuss its relationship with the generating functional of the Chern-Simons theory. 

 \subsection{Connected diagrams}
 
 In the computation of  the path integral  which appears  in the numerator of expression (\ref{2.15}), it is convenient to make the linear change of variables 
\be
A_\mu^a \longrightarrow A_\mu^a + \widehat A_\mu^a \quad , \quad B_\mu^a \longrightarrow B_\mu^a + \widehat B_\mu^a \quad \; , 
\label{3.1}
\ee
in which $A^a_\mu $ and $B^a_\mu$ are called the quantum components, whereas the classical components $\widehat A_\mu^a$ and $\widehat B_\mu^a$ satisfy the equations of motion in the presence of the sources 
\be
{\delta S[\widehat A , \widehat B ]\over \delta \widehat B^a_\mu (x)} + K^a_\mu(x) = 0 \quad , \quad {\delta S[\widehat A , \widehat B ]\over \delta \widehat A^a_\mu (x)} + J^a_\mu(x) = 0 \; , 
\label{3.2}
\ee
together with the gauge-fixing constraints 
\be
\der^\mu \widehat A^a_\mu (x) = 0 = \der^\mu \widehat B^a_\mu (x) \;  . 
\label{3.3}
\ee
Because of equations (\ref{3.2}), the classical components $\widehat A_\mu^a$ and $\widehat B_\mu^a$ are functions of $J^a_\mu $ and $K^a_\mu$, (and, for localised  $J^a_\mu $ and $K^a_\mu$, both components  $\widehat A_\mu^a$ and $\widehat B_\mu^a$ vanish in the $|x| \rightarrow \infty $ limit as $\sim 1/ |x|^2$).  One then finds 
\bea
&&{\hskip -1 cm}S_{TOT} [A + \widehat A , B + \widehat B,...] + \int d^3x \left [ J^a_\mu (A^a_\mu + \widehat A^a_\mu ) + K^a_\mu (B^a_\mu + \widehat B^a_\mu ) \right ] = \nonumber \\
&& {\hskip 2 cm} = S[\widehat A , \widehat B] 
+ \int d^3x \left [ J^a_\mu  \widehat A^a_\mu + K^a_\mu  \widehat B^a_\mu  \right ] + \widetilde S[A, B, ...] \; , 
\label{3.4}
\eea
where 
\bea
\widetilde S [A,B,...] &=& S_{TOT}[A,B,M,N,\xi, \overline \xi, c , \overline c] \nonumber \\
&& {\hskip - 0.5 cm} - \int d^3x \, \epsilon^{\mu \nu \lambda} \epsilon^{abc} \left [  B^a_\mu \widehat A^b_\nu A^c_\lambda + \mezzo \widehat B_\mu^a A^b_\nu A^c_\lambda + \mezzo g A^a_\mu \widehat A^b_\nu A^c_\lambda \right ] \nonumber \\ 
&&- \int d^3x \, \epsilon^{abc} \left [ \der^\mu \overline c \, \widehat A^b_\mu c^c + \der^\mu  \overline \xi \, \widehat A^b_\mu   \xi^c + \der^\mu \overline \xi^a \, \widehat B^b_\mu c^c \right ] 
\; . 
\label{3.5}
\eea
Note that  $\widetilde S [A, B,... ]$ represents the resulting    action for the quantum components $A^a_\mu$ and $B^a_\mu$ of the fields in which 

\begin{itemize}
\item the linear terms in the quantum fields are  missing. Indeed,  as a consequence of equations (\ref{3.2}) and (3.3), $\widehat A_\mu^a$ and $\widehat B_\mu^a$ satisfy the classical gauge-fixing constraint and represent a stationary point of the action in the presence of the source terms; 

\item the lagrangian vertices for the quantum fields ---which are contained in $\widetilde S[A,B,...]$--- depend on the $J^a_\mu$ and $K^a_\mu$ through the classical components  $\widehat A^a_\mu$ and $\widehat B^a_\mu$. 

\end{itemize}

\noindent Therefore the generating functional $W[J,K]$ satisfies 
\be
e^{i W[J,K]}  = e^{i S[\widehat A , \widehat B] 
+ i \int d^3x \left [ J^a_\mu  \widehat A^a_\mu + K^a_\mu  \widehat B^a_\mu  \right ] }
{\int D\hbox{(fields) } e^{i \widetilde S } \over 
\int D\hbox{(fields) } e^{iS_{TOT}}  } \; .
\label{3.6}
\ee
This expression  shows  that  $W[J,K]$ can be written as the sum of two parts, $W=W_0 + W_1$, in which  

\begin{itemize}
\item the connected tree-level Feynman diagrams entering $W_0 $  are described by a Legendre transformation of the classical action,  
\be
W_0[J,K] = S[\widehat A , \widehat B] 
+  \int d^3x \left [ J^a_\mu  \widehat A^a_\mu + K^a_\mu  \widehat B^a_\mu  \right ] \; ; 
\label{3.7}
\ee
\item  the connected diagrams containing loops ---described by $W_1$---   are obtained by computing the vacuum-to-vacuum diagrams of the quantum field components. These diagrams  are determined  by the lagrangian terms contained in  the resulting action $\widetilde S$, with the normalization given by the vacuum-to-vacuum diagrams computed in the absence of sources, {\it i.e.}, when $\widehat A^a_\mu$ and  $\widehat B^a_\mu$ vanish.
\end{itemize}

\smallskip

\begin{prop}
The function $W_0[J,K]$  is given by 
\be
W_0[J,K] = g \, S_{CS} [\widehat A \, ] + \int d^3x \, J^a_\mu (x) \widehat A^a_\mu (x) \; , 
\label{3.8}
\ee
where the Chern-Simons action $S_{CS} [A]$ is shown in equation {\rm (\ref{2.9})}; $\widehat A^a_\mu $ is a classical field  which only depends on $K^a_\mu$, it satisfies   $\der^\mu \widehat A^a_\mu (x) =0$ and 
\be
{\der S_{CS} [\widehat A \, ] \over \delta \widehat A^a_\mu (x)} = - K^a_\mu (x)\; . 
\label{3.9}
\ee

\end{prop}

\smallskip

\noindent {\bf Proof.}  Since the BF action (\ref{2.8})  can be written as 
\be
S[A,B] =  \int d^3 x \,  B^a_\mu (x){\delta S_{CS}[A] \over \delta A^a_\mu(x) } + g \, S_{CS}[A]  \; , 
\label{3.10}
\ee  
the first of equations (\ref{3.2}) coincides with equation (\ref{3.9}). This means that $\widehat A^a_\mu (x)$ only depends on $K^a_\mu $ and does not depend on $J^a_\mu$ and  $g$. Finally, the action $S[A,B]$ is  a linear function of $B^a_\mu$.  Therefore, in the Legendre transform (\ref{3.7}),  the two terms which are linear in $\widehat B^a_\mu$ cancel, and one obtains  precisely expression (\ref{3.8}). {\hfill \cvd}

\medskip

Equation (\ref{3.8}) is in agreement with expression (\ref{2.17}), and shows that when $K^a_\mu  = (1/g) J^a_\mu$, the functional   $W_0[J,(1/g)J]$ satisfies 
\be
W_0[J,(1/g)J] = W_{0, CS} [J] \; , 
\label{3.11}
\ee
where $W_{0, CS} [J] $ denotes the generating functional  of the tree-level connected diagrams of the Chern-Simons theory, which is defined by the action $g S_{CS}[A]$,  
\be
W_{0, CS} [J] = g \, S_{CS} [\widehat A] + \int d^3x \, J^a_\mu (x) \widehat A^a_\mu (x) \; , \quad \hbox{with  }~~  g{\der S_{CS} [\widehat A \, ] \over \delta \widehat A^a_\mu (x)}= - J^a_\mu (x)\; . 
\label{3.12}
\ee

Let us now consider diagrams with loops. 

\smallskip 

\begin{prop}
The whole set  of the  vacuum-to-vacuum connected diagrams for the quantum field components is equal to the set $\, i W_1[K] $ of  the one-loop  connected diagrams which only depend on $K_\mu^a$, 
 \be
   e^{i W_1[K]}  = \langle e^{-i  \int d^3x   \epsilon^{abc}  \{ \epsilon^{\mu \nu \lambda} B^a_\mu  \widehat A^b_\nu A^c_\lambda +  \der^\mu \overline c^a \, \widehat A^b_\mu c^c +  \der^\mu  \overline \xi^a \, \widehat A^b_\mu   \xi^c   \} }\rangle \; . 
 \label{3.13}
 \ee

\end{prop}

\smallskip 

\noindent {\bf Proof.} The field propagators that are derived from the $S_{TOT}$ are shown in equations (\ref{2.12}) and (\ref{2.13}); in particular,  it turns out that $\WT{A^a_\mu (x)A}\null^b_\nu (y) =0$ and $\WT{\xi (x) \overline c}(y) = 0 = \WT{c (x)\overline \xi}(y)$. Consequently, the only connected source-dependent diagrams containing loops  are the one-loop  connected diagrams  entering  equation (\ref{3.13}). {\hfill \cvd} 

\medskip

The result (\ref{3.13}) is in agreement with the statements of Proposition~1  and shows that,  when $K^a_\mu = (1/g) J^a_\mu $,  the functional $W_1[(1/g) J ]$ verifies 
\be
W_1[(1/g) J ] = 2 W_{1, CS}[J] \; , 
\label{3.14}
\ee
where the factor 2 is due to the combinatorics and the presence of two ghost fields, and $W_{1, CS}[J]$ denotes  the  generating functional of the one-loop connected diagrams in the Chern-Simons theory, 
\be
 e^{i W_{1, CS}[J]}   = \langle e^{-i  \int d^3x   \epsilon^{abc}  \{ (g/2) \epsilon^{\mu \nu \lambda}   A^a_\mu  \widehat A^b_\nu A^c_\lambda +  \der^\mu \overline c \, \widehat A^b_\mu c^c    \} } \, \rangle \Big |_{CS} \; . 
 \label{3.15}
 \ee  
 
 \subsection{Connected one-loop diagrams}
 
 As a consequence of equation \eqref{3.13}, the functional $ W_1[K]$ can be written as 
  \be
 W_1 [K ] = W_1^{(v)} [K] + W_1^{(g)} [K] \; , 
 \label{3.16}
 \ee
 where $W_1^{(v)} [K] $ corresponds to  the sum of the connected diagrams with one loop of the vector fields, whereas $W_1^{(g)} [K]$  denotes the sum of the connected diagrams with one loop of the ghost fields. 
In Schwinger notations \cite{SCW}, the $\WT{AB}$ propagator \eqref{2.12} reads  
 \be
 \WT{A^a_\mu (x) B}\! \null^b_\nu(y) = \langle x \, ;  a , \mu  |  \,  i \frac{ \epsilon_{\mu \lambda \nu} \der^\lambda  }{\der^2}  \,  | y  \, ; b , \nu  \rangle \; , 
 \label{3.17}
  \ee
 and then 
 \bea
i W_1^{(v)}[K] 
 &=&  \sum_{n=1}^\infty \frac{i^n}{n} \, \left[ \sum \int d^3 x_1 \ldots d^3 x_{n} \, \,  \langle {x_1} |    \WT{A \> B} \, \widehat A \, | {x_2} \rangle  \cdots \langle {x_{n}} |  \WT{A \> B} \, \widehat A \, | {x_1} \rangle \right ] \nonumber \\
 &=&   \sum_{n=1}^\infty \, \frac{(-1)^n}{n}\, \hbox{Tr} \left [ \frac{1}{\der^2}   \,  \widehat M  \, \right ]^n    
  \; , 
 \label{3.18}
 \eea
where 
\be
\left (      \widehat M \right )^{a,c\, ; \, \nu}_{ ~ ~ ~   \mu}  = \epsilon_{\mu \tau \sigma} \, \der^\tau \epsilon^{a  b c} \epsilon^{\sigma \lambda  \nu} \widehat A^b_\lambda \; ,  
\label{3.19}
\ee
 and $\hbox{Tr}$ denotes the trace in the color indices, vector indices  and orbital indices 
  \be
\hbox{Tr } (Q) = \sum_{a , \mu }\int d^3 x \,  \langle x \, ; a , \mu  | \, Q \,  | x \, ; a , \mu \rangle  \; .  
\label{3.20}
\ee
 The connected diagrams with one loop of the ghost fields give the contribution 
 \be
i W_1^{(g)}[K]   = - 2 \sum_{n=1}^\infty \, \frac{1}{n}\, \hbox{tr} \left [ \frac{1}{\der^2}\,  \widehat N  \, \right ]^n      \; , 
 \label{3.21}
 \ee
in which 
 \be
\left (   \widehat N \right )^{a,c}  =  \der^\lambda \,  \epsilon^{a  b c} \widehat A^b_\lambda \; ,  
\label{3.22}
\ee
and 
  \be
\hbox{tr } (Q) = \sum_{a }\int d^3 x \,  \langle x \, ; a   | \, Q \,  | x \, ; a  \rangle  \; .  
\label{3.23}
\ee
Some diagrams contained in $W_1 [K]$ have values which are not well defined because of possible ultraviolet divergences; these diagrams  will be  renormalized in the Section~4. 

\section{Renormalization}

Since the observables that we shall consider  only depend on  $A^a_\mu$ and $B^a_\mu$,  and since the corresponding BF connected diagrams have zero loops or one loop only,  in order to discuss the relevant aspects of the renormalization we need to consider the functional 
\be
\Gamma = S_{TOT}[A,B,M,N,\xi, \overline \xi, c , \overline c] + \Gamma_1 [A,B] \; , 
\label{4.1}
\ee
where $i \Gamma_1 [A,B]$ denotes to the sum of the one-particle-irreducible diagrams with  one loop,   in which $A^a_\mu$ and $B^a_\mu $ represent the  external legs \cite{BO,PS,IZ}. In other words, $\Gamma_1 [A,B]$ is the sum of the one-loop   proper vertices  for the field components $A^a_\mu$ and $B^a_\mu $. Indeed, as it has been shown in Section~2  and in Section~3, in the BF theory the contributions to the proper vertices which are described by diagrams with two or more loops are absent. The zero-loop component of the proper vertices coincides with the lagrangian and the one-loop component  only contains primitive divergences. Therefore, 
 in the renormalization procedure,  diagrams with external ghost fields can be ignored. 

Equations (\ref{2.18}), (\ref{3.18}) and (\ref{3.21}) imply that  $ \Gamma_1[A,B]$ nontrivially depends  on $A^a_\mu$ only, 
\be
\Gamma_1[A,B] = \Gamma_1[A] \; . 
\label{4.2}
\ee
Each term of the expansion of $\Gamma_1[A]$ in powers of the fields $A^a_\mu $  is well defined apart from the terms with two and three fields. The corresponding six diagrams are not well defined a priori; they possibly have ultraviolet divergences. Since  only a finite number of diagrams need to be renormalized,  the BF model is a superrenormalizable field theory. 

\subsection{Normalization conditions}
As there are no gauge anomalies in three  dimensions, it is possible to define a renormalized $\Gamma$ which is BRST invariant. Let us define 
\be
{\delta^2 \Gamma \over \delta A^b_\nu (y) \delta A^a_\mu (x)} \Bigg |_{A=0  , B=0} =  
\int  {d^3k \over (2 \pi )^3} e^{i k (x-y)} \Pi^{ab}_{\mu \nu} (k) \; ,    
\label{4.3}
\ee
\be
{\delta^2 \Gamma \over \delta A^b_\nu (y) \delta B^a_\mu (x)} \Bigg |_{A=0  , B=0} =  
\int  {d^3k \over (2 \pi )^3} e^{i k (x-y)} \Sigma^{ab}_{\mu \nu} (k) \; .    
\label{4.4}
\ee
In addition to the BRST invariance of $\Gamma$,  the normalization conditions are taken to be 
\be
\lim_{k \rightarrow 0} \Pi^{ab}_{\mu \nu} (k) = i g \delta^{ab} \epsilon^{\mu \lambda \nu } k_\lambda \; ,   
\label{4.5}
\ee
and
\be
\lim_{k \rightarrow 0} \Sigma^{ab}_{\mu \nu} (k) = i  \delta^{ab} \epsilon^{\mu \lambda \nu } k_\lambda \; .  
\label{4.6}
\ee
Equations (\ref{4.5}) and (\ref{4.6}) establish  the normalization of the fields and specify  the value of the coupling constant $g$. Since the one-loop contributions contained in $\Gamma_1[A]$ do not depend on the field $B^a_\mu$, equation (\ref{4.6})   ---which is valid at the tree-level--- remains valid to all orders of perturbations theory. Consequently,  only equation (\ref{4.5}) needs to be considered; in renormalized perturbation theory \cite{PS}, equation (\ref{4.5}) controls the  one-loop counterterms.    Let us consider the renormalization procedure  \cite{BO,HL,LAZ,GB,KAI,NST} in the space of the coordinates $x^\mu$.  Of course, the final result coincides with the result obtained by means of the  renormalization procedure in momentum space. 

\subsection{One-loop two points function}

$\Gamma_1 [A]$ can be expanded in powers of the fields $A_\mu^a$; the quadratic term   is given by the sum of the contribution  $ \Gamma^{(v)}_1 [A]$, corresponding to the one-loop diagram of Figure~4(a), and $ \Gamma^{(g)}_1 [A]$ which is obtained by adding the two equal amplitudes which are described by the diagram of Figure~4(b) containing one loop of the two types of ghosts. 

\vskip 0.7 truecm

\centerline {
\begin{tikzpicture} [scale=0.9] [>=latex]
\draw [very thick  ] (0,0) circle (0.8); 
\draw [very thick]  (-2,0) -- (-0.8,0); 
\draw [very thick]  (0.8,0) -- (2,0);
\draw [very thick , dashed ] (7,0) circle (0.8); 
\draw [very thick] (5,0) -- (6.2,0); 
\draw [very thick] (7.8,0) -- (9,0); 
\node at (0,-1.8) {(a)};
\node at (7,-1.8) {(b)};
\end{tikzpicture}
}

\vskip 0.4 truecm
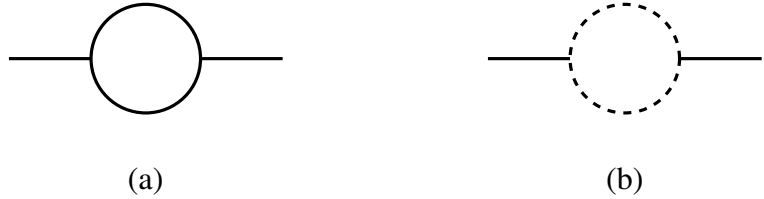
\centerline {{Figure 4.} {One loop contributions to the two points function.}}

\vskip 0.7 truecm

\noindent One has 
\bea
i \Gamma_1^{(v)}[A] &=& {(-i)^2 \over 2 !} \! \! \int d^3x \, d^3y \, A^a_\mu (x) A^b_\nu (y) \epsilon^{cad} \epsilon^{ebh} 
 \epsilon^{\lambda \mu \tau } \epsilon^{\sigma \nu \alpha} \WT{A^d_\tau (x) B}\! \null^e_\sigma(y)\, \WT{A^h_\alpha (y) B}\! \null^c_\lambda (x)\nonumber \\ 
 &=& -2 \int d^3x \, d^3y \, A^a_\mu (x) A^b_\nu (y)\, \delta^{ab} \, \der^\mu_x  \Delta(x-y) \, \der^\nu_y \Delta(y-x) \; , 
 \label{4.7}
\eea
and 
\bea
i \Gamma_1^{(g)}[A] &=& - (-i)^2  \! \! \int d^3x \, d^3y \, A^a_\mu (x) A^b_\nu (y) \epsilon^{cad} \epsilon^{ebh}   \WT{c^d (x) \der^\nu {\overline c}}\! \null^e (y)\, \WT{c^h (y) \der^\mu {\overline c}}\! \null^c (x)\nonumber \\ 
 &=& 2 \int d^3x \, d^3y \, A^a_\mu (x) A^b_\nu (y)\, \delta^{ab} \, \der^\mu_x  \Delta(x-y) \, \der^\nu_y \Delta(y-x) \; .  
 \label{4.8}
\eea
Precisely like  in  the Chern-Simons theory \cite{GMMO, ES}, the sum of the two contributions $\Gamma^{(v)}_1[A] + \Gamma^{(g)}_1[A]$ formally vanishes, indeed 
\be
\Gamma^{(v)}_1[A] + \Gamma^{(g)}_1[A] = 2i  \int d^3x \, d^3y \, A^a_\mu (x) A^b_\nu (y)\, \delta^{ab} \, H^{\mu \nu} (x,y) \; , 
\label{4.9}
\ee
where 
\be
H^{\mu \nu} (x,y) = \der^\mu_x  \Delta(x-y) \, \der^\nu_y \Delta(y-x)  - \der^\mu_x  \Delta(x-y) \, \der^\nu_y \Delta(y-x)  \; . 
\label{4.10}
\ee
The  amplitude  
\be
\der^\mu_x  \Delta(x-y) \, \der^\nu_y \Delta(y-x) =  { (x-y)^\mu (y -x)^\nu \over (4 \pi)^2 \, |x-y |^6} \; , 
\label{4.11} 
\ee
 which appears in equation (\ref{4.10}), is well defined for $x\not=y$.   Consequently ``the nonlocal component'' of $\Gamma^{(v)}_1[A] + \Gamma^{(g)}_1[A]$ is well defined and  vanishes because   
\be
H^{\mu \nu} (x,y) \Big |_{x\not= y } = 0 \; . 
\label{4.12}
\ee 
When $x=y$ expression (\ref{4.11}) is not well defined, so one has to specify the value of $H^{\mu \nu} (x,y)$ in the case   $x=y$.   In facts, since  ``the nonlocal component'' of $\Gamma^{(v)}_1[A] + \Gamma^{(g)}_1[A]$  vanishes,  the entire renormalization of $\Gamma^{(v)}_1[A] + \Gamma^{(g)}_1[A]$ consists \cite{BO} precisely in specifying  the value of ``the local component'' of $\Gamma^{(v)}_1[A] + \Gamma^{(g)}_1[A]$, which is defined by $H^{\mu \nu} (x,y)$ for $x=y$. This value is uniquely determined by the normalization condition (\ref{4.5}), which requires
\be
\left  ( \Gamma^{(v)}_1[A] + \Gamma^{(g)}_1[A] \right ) \Big |_{\rm renormalized} = 0 \; . 
\label{4.13}
\ee    

It should be noted that the renormalized value (\ref{4.13}) of $\Gamma^{(v)}_1[A] + \Gamma^{(g)}_1[A]$ is also in agreement with  the point-splitting procedure, that we shall use  in the definition of the composite Wilson line operators. Indeed,  the point-splitting definition of $H^{\mu \nu} (x,y)$ for $x=y$  gives   
\be
H^{\mu \nu} (x,y) \Big |_{x = y } \equiv \lim_{x \rightarrow y} H^{\mu \nu} (x,y) \Big |_{x\not= y } = 0 \; ,  
\label{4.14}
\ee
which implies precisely equation (\ref{4.13}). 
  
From equation (\ref{4.13}) it follows  that  the BF vacuum polarisation vanishes and the Feynman propagators (\ref{2.12}) coincide with the dressed propagators. 

\subsection{One-loop three points function}

The term of $\Gamma_1 [A]$ which contains three powers of the field $A^a_\mu $ is the sum of $\widetilde \Gamma^{(v)}_1 [A]$, which is described by the Feynman diagram of Figure~5(a), and $\widetilde \Gamma^{(g)}_1 [A]$ 
which is specified by the one-loop contributions of Figure~5(b)  induced by the two kinds of ghosts.

\vskip 0.7 truecm

\centerline {
\begin{tikzpicture} [scale=0.9] [>=latex]
\draw [very thick  ] (0,0) circle (0.8); 
\draw [very thick]  (-2,0) -- (-0.8,0); 
\draw [very thick]  (0.8,0) -- (2,0);
\draw [very thick]  (0,0.8) -- (0,2);
\draw [very thick , dashed ] (7,0) circle (0.8); 
\draw [very thick] (5,0) -- (6.2,0); 
\draw [very thick] (7.8,0) -- (9,0); 
\draw [very thick]  (7,0.8) -- (7,2);
\node at (0,-1.8) {(a)};
\node at (7,-1.8) {(b)};
\end{tikzpicture}
}

\vskip 0.4 truecm
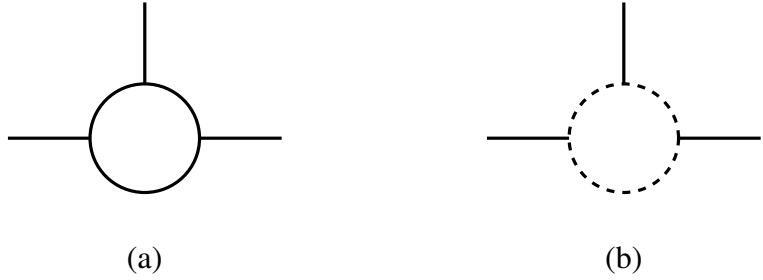
\centerline {{Figure 5.} {One loop contributions to the three points function.}}

\vskip 0.7 truecm

\noindent One finds 
\bea
i \widetilde \Gamma^{(v)}_1[A]  &=&  {2 (-i)^3 \over 3!} \! \! \int d^3x \, d^3y \, d^3z \, A^b_\nu (x) A^e_\rho (y) A^h_\tau (z) \, \epsilon^{abc} \epsilon^{def} \epsilon^{ghi} \nonumber \\
&& \qquad \epsilon^{\mu \nu \lambda} \epsilon^{\sigma \rho \gamma} \epsilon^{\alpha \tau \beta} \WT{A^c_\lambda (x) B}\! \null^d_\sigma(y)\, \WT{A^f_\gamma (y) B}\! \null^g_\alpha (z)\, \WT{A^i_\beta (z) B}\! \null^a_\mu (x)\nonumber \\
&=& {1\over 3} \int d^3x \, d^3y \, d^3z \, \epsilon^{abc} \, A^a_\mu (x) A^b_\nu (y) A^c_\lambda (z) \, T^{\mu \nu \lambda}_{\tau \rho \sigma} \nonumber \\
 && \qquad \der_x^\tau \Delta (x-y) \, \der_y^\rho \Delta (y-z) \, \der_z^\sigma \Delta (z-x) \; , 
 \label{4.15}
\eea
where
\be
T^{\mu \nu \lambda}_{\tau \rho \sigma} = \delta^\mu_\tau \delta^\nu_\sigma  \delta^\lambda_\rho  
+  \delta^\mu_\sigma \delta^\nu_\rho  \delta^\lambda_\tau +  \delta^\mu_\rho \delta^\nu_\tau  \delta^\lambda_\sigma -  \delta^\mu_\rho \delta^\nu_\sigma  \delta^\lambda_\tau \; . 
\label{4.16}
\ee
Moreover 
\bea
i \widetilde \Gamma^{(g)}_1[A]  &=&  - 4 {(-i)^3 \over 3!}  \! \! \int d^3x \, d^3y \, d^3z \, A^b_\mu (x) A^e_\nu (y) A^h_\lambda (z) \, \epsilon^{abc} \epsilon^{def} \epsilon^{ghi} \nonumber \\
&& \qquad \WT{c^c (x) \der^\nu {\overline c}}\! \null^d (y)\, \WT{c^f (y) \der^\lambda {\overline c}}\! \null^g (z)\, \WT{c^i (z) \der^\mu {\overline c}}\! \null^a (x)\nonumber \\
&=& -  {1\over 3} \int d^3x \, d^3y \, d^3z \, \epsilon^{abc} \, A^a_\mu (x) A^b_\nu (y) A^c_\lambda (z) ( \delta^\mu_\sigma \delta^\nu_\tau   \delta^\lambda_\rho  +  \delta^\mu_\tau \delta^\nu_\rho  \delta^\lambda_\sigma  )   \nonumber \\
 && \qquad \; \der_x^\tau \Delta (x-y) \, \der_y^\rho \Delta (y-z) \, \der_z^\sigma \Delta (z-x)  \; . 
 \label{4.17}
\eea
Therefore 
\be
i \widetilde \Gamma^{(v)}_1[A]  +  i \widetilde \Gamma^{(g)}_1[A]   =  {1\over 3} \int d^3x \, d^3y \, d^3z \, \epsilon^{abc} \, A^a_\mu (x) A^b_\nu (y) A^c_\lambda (z) \, V^{\mu \nu \lambda } (x,y,z) \; , 
 \label{4.18}
\ee
in which 
\be
V^{\mu \nu \lambda } (x,y,z)  =  \epsilon^{\mu \nu \lambda}_{\tau \rho \sigma} \, \der_x^\tau \Delta (x-y) \, \der_y^\rho \Delta (y-z) \, \der_z^\sigma \Delta (z-x) \; , 
 \label{4.19}
\ee
and 
\be
\epsilon^{\mu \nu \lambda}_{\tau \rho \sigma} = 
 \delta^\mu_\tau \delta^\nu_\sigma  \delta^\lambda_\rho  
+  \delta^\mu_\sigma \delta^\nu_\rho  \delta^\lambda_\tau +  \delta^\mu_\rho \delta^\nu_\tau  \delta^\lambda_\sigma -  \delta^\mu_\rho \delta^\nu_\sigma  \delta^\lambda_\tau - \delta^\mu_\sigma \delta^\nu_\tau   \delta^\lambda_\rho  -  \delta^\mu_\tau \delta^\nu_\rho  \delta^\lambda_\sigma \; . 
\label{4.20}
\ee
When $x\not= y $, $ x \not=z$ and $y \not= z$, the amplitude 
\be
\der_x^\tau \Delta (x-y) \, \der_y^\rho \Delta (y-z) \, \der_z^\sigma \Delta (z-x) = { (x-y)^\tau (y-z)^\rho (x-z)^\sigma  \over (4 \pi )^3 \, |x-y|^3 |y-z|^3 |z-x|^3} 
\label{4.21}
\ee
is well defined and,  when it is multiplied by the completely antisymmetric tensor $\epsilon^{\mu \nu \lambda}_{\tau \rho \sigma}$, it vanishes, 
\be
V^{\mu \nu \lambda } (x,y,z) \Big |_{x\not= y \not= z} = 0 \; . 
\label{4.22}
\ee 
Therefore, as in the case of the two points functions, ``the nonlocal component'' of $\widetilde \Gamma^{(v)}_1[A]  +   \widetilde \Gamma^{(g)}_1[A] $  is vanishing.  In order to specify  the renormalized value of $\widetilde \Gamma^{(v)}_1[A]  +   \widetilde \Gamma^{(g)}_1[A]$ we need to define \cite{HL,LAZ,GB,KAI,NST}  the value of the ``diagonal local component'' of $V^{\mu \nu \lambda } (x,y,z) $, corresponding to the case in which the external fields are defined at coincident points $x=y=z$. This is in agreement with the general fact that, in one-loop diagrams, the possibly divergent (not well defined) contribution is local or, to be more precise, the introduction of  appropriate local counterterms makes the diagrams well defined.  

The renormalized value of $\widetilde \Gamma^{(v)}_1[A]  +   \widetilde \Gamma^{(g)}_1[A]$ is determined by the normalization conditions and by symmetry arguments.  Indeed the  BRS invariance of $\Gamma $ requires that the value of the  local component of the one-loop contribution to the   3-point proper vertex  must be $(1/6)$ the value of the one-loop contribution to the dressed propagator, which vanishes. Therefore relation (\ref{4.13}) and BRST invariance  imply 
\be
\left ( \widetilde \Gamma^{(v)}_1[A]  +   \widetilde \Gamma^{(g)}_1[A] \right ) \Big |_{\rm renormalized} = 0 \; . 
\label{4.23}
\ee    
The result (\ref{4.23}) can  also be obtained 
 by means of the point-splitting procedure, according to which  
\be
V^{\mu \nu \lambda } (x,y,z) \Big |_{x= y = z} = \lim_{x \rightarrow y}\,  \lim_{y\rightarrow z} \, V^{\mu \nu \lambda } (x,y,z) \Big |_{x\not= y \not= z} = 0 \; . 
\label{4.24}
\ee
The point-splitting procedure also shows that each ``partially local component'', say $ x= y \not= z$, is vanishing because 
$$
V^{\mu \nu \lambda } (x , y , z) \Big |_{x = y \not= z} = \lim_{x \rightarrow y}\,   V^{\mu \nu \lambda } (x,y,z) \Big |_{x\not= y \not= z} = 0 \; . 
$$
In  renormalizable field theories, the ``partially local components'' of the diagrams are possibly related with  the (overlapping) sub-divergences. In the connected diagrams of the BF theory, there are no subdivergences   to deal with  because the connected diagrams have  at most one loop.   
 
Since all the remaining diagrams contributing to $\Gamma$ are finite, this concludes the renormalization of the BF theory in ${\mathbb R}^3$. This means that, by taking into account equations (\ref{4.13}) and (\ref{4.23}), the expectation values  
\be
 \langle A_{\mu_1}^{a_1}(x_1)A_{\mu_2}^{a_2}(x_2)\cdots A^{a_n}_{\mu_n} (x_n) B^{c_1}_{\nu_1} (y_1) B^{c_2}_{\nu_2} (y_2) \cdots B^{c_m}_{\nu_m}(y_m) \rangle \; , 
 \label{4.25}
 \ee
when the fields are defined at not coincident points,  are well defined.   In the computation of the BF   observables,  we shall need to remove certain ambiguities of the expectation values  which appear in a specific limit in which  two fields are defined in the same point. This issue, which is related to the introduction of a framing for the knots,  will be discussed in Section~7.  

\section {Wilson line observables}

Similarly to the case of the Chern-Simons gauge field theory, the gauge invariant observables that we shall consider correspond to  appropriately normalized  traces of the 
expectation values of  the gauge holonomies which are associated with oriented framed knots   in ${\mathbb R}^3$ in a given representation of $ISU(2)$. 

\subsection{Representations of $\ner {ISU(2)}$}
We shall consider  linear unitary representations of $ISU(2)$ in which $\{ P^a \}$ are nontrivially represented  and which are  specified by the values of the two quadratic Casimir operators $P^a P^a$ and $J^a P^a$.  More precisely,  if  $| \varphi \rangle $ denotes a vector transforming according to  the  irreducible $(\Lambda , r )$ representation, it must satisfy   
\be
P^a P^a | \varphi \rangle  = \Lambda^2 | \varphi \rangle \; , 
\label{5.1}
\ee
and 
\be
J^a P^a | \varphi \rangle = r \Lambda | \varphi \rangle \; , 
\label{5.2}
\ee
with fixed positive $\Lambda $ and fixed semi-integer $r$ ({\it i.e.}, $2 r \in {\mathbb Z}$). In  this article we shall  concentrate on the ``scalar" $(\Lambda , 0 )$ representation and  the ``fundamental" $(\Lambda , 1/2)$ representation. 

In order to describe these representations, let us first consider the  quantum mechanics  states space of a spinless  particle moving in three dimensional euclidean space. Let $P^a$ represent the cartesian components of the  momentum operator and  let $L^a$  denote the components of the orbital angular momentum of the particle, 
 \be
 L^a = \epsilon^{abc } Q ^b P^c    \; ,  
 \label{5.3} 
 \ee
 in which $[Q^a , P^b ] = i \delta^{ab} $.   The operators $\{ J^a = L^a , P^a \}$ satisfy the commutation relations (\ref{2.2}). 
 
 \subsubsection{Scalar representation}
 
 The plane wave  
\be
\psi_{\bm k} (\bm r) = e^{i \bm k \bm r} 
\label{5.4}
\ee
verifies  
\be
P^a \, \psi_{\bm k} (\bm r) = k^a \,  \psi_{\bm k} (\bm r) \; . 
\label{5.5}
\ee
 When the value of the first Casimir operator $P^a P^a$ of $ISU(2)$  is chosen to be $\Lambda^2$, one needs to consider  the linear space ${\cal H}_{(\Lambda , 0 )}$ which is generated by all the vectors 
 \be
  \{ \psi_{\bm k} (\bm r)  \} \hbox{ with } \ner k \ner k = \Lambda^2 \; . 
  \label{5.6}
 \ee
 In this case, the  possible eigenvalues $\ner k$ of the momentum belong to  a spherical surface  in ${\mathbb R}^3$ of radius equal to $\Lambda$.  The set of all the plane waves $\{ \psi_{\bm k} (\bm r) \}$  with $\bm k \bm k = \Lambda^2$  is left invariant by the action of the $SU(2)$ group which is generated by the orbital angular momentum components \eqref{5.3}.  Therefore the linear  space ${\cal H}_{(\Lambda , 0 )}$ is invariant under  the  transformations   generated by  $\{  J^a = L^a, P^a  \} $.   Since $L^a P^a =0$, the $ISU(2)$ action on ${\cal H}_{(\Lambda , 0 )}$  which is implemented by the transformations $ \exp \left \{  i   \beta^a J^a + i \eta^a P^a \right \} $  defines  the scalar  $(\Lambda , 0 )$ representation of $ISU(2)$. 
 
The commutation relations of the  operators $\{  J^a = L^a, Q^a  \} $ also coincide with the commutation relations of the $ISU(2)$ algebra. Thus an alternative interpretation of this $ISU(2)$ representation can be obtained by considering the quantum mechanics states of one particle moving on the surface of a 2-sphere in ${\mathbb R}^3$.  For the purposes of the present article, we don't need to discuss the rigged Hilbert space structure \cite{RH} associated with ${\cal H}_{(\Lambda , 0 )}$.  

\subsubsection{Fundamental representation}

Let us now examine the fundamental $(\Lambda , 1/2)$ representation of $ISU(2)$. Let $ {\cal H}_{spin}$ denote the two dimensional space of the  spin states of a spin $(1/2)$ nonrelativistic particle, and let  $S^a$  represent the components of the spin operator,    
 \be
 S^a = \mezzo \sigma^a \; , 
 \label{5.7}
 \ee
 where $\sigma^a $ denote the Pauli sigma matrices. The operators $S^a$  act on the vectors which belong to ${\cal H}_{spin}$.  In the tensor product ${\cal H}_{(\Lambda , 0 )}  \otimes {\cal H}_{spin}$,  one can put 
\be
J^a = L^a + S^a \; . 
\label {5.8}
 \ee
 In addition to the constraint $\ner k \ner k = \Lambda^2$,  the specification of the value $(1/2) \Lambda$ of the second Casimir operator $J^a P^a$ selects the states 
 in ${\cal H}_{(\Lambda , 0 )}   \otimes {\cal H}_{spin}$ 
 of positive helicity. Let $\pi_+$ denote the projector on the positive helicity states,  
\be
\pi_+ = \mezzo \left ( 1 + {\ner P \ner \sigma \over \Lambda } \right ) \; . 
\label{5.9}
\ee
Let ${\cal H}_{(\Lambda, 1/2)}$ be the linear space which is generated by the vectors   
\be
\bigl \{ \, \pi_+ \, |  \chi \rangle \, \bigr \} \hbox { in which }  \, |  \chi \rangle \in {\cal H}_{(\Lambda , 0 )}  \otimes {\cal H}_{spin}  \; . 
\label{5.10}
\ee
The $ISU(2)$ action on ${\cal H}_{(\Lambda , 1/2 )} $,  which is carried out by the transformations generated by  $\{  J^a = L^a + S^a , P^a  \} $, defines  the $(\Lambda, 1/2)$ representation. One can easily verify that the projector  $\pi_+$ commutes with the generators of $ISU(2)$. 

A generic $(\Lambda , r)$ representation could  be constructed by means of a  procedure which is similar to the procedure that has been illustrated in the case of the $(\Lambda , 1/2)$ representation.  Each representation 
$(\Lambda , r)$, with $r=0$ or $r=1/2$,   is irreducible and infinite dimensional.  
 
 \subsection{Holonomies} 
 
 Let us consider a classical gauge configuration which is described by the components $A^a_\mu(x)$ and $ B^a_\mu (x)$.  Given  an oriented path $\gamma $  in ${\mathbb R}^3$, which connects the starting point $x_1$ to the final point $x_2$, the corresponding $ISU(2)$ holonomy $h_\gamma \in ISU(2)$   is  defined by 
 \be
h_\gamma =  {\rm P} e^{i \int_\gamma  d x^\mu (A^a_\mu (x) J^a + B^a_\mu (x) P^a )  } \; , 
\label{5.11}
 \ee
 where the symbol $\rm P$ denotes the path-ordering of the $\{J^a, P^b \} $ operators along the direction specified by the orientation of $\gamma$.    Under a gauge transformation (\ref{2.3}), $h_\gamma$  transforms as 
 \be
 h_\gamma \rightarrow \Omega^{-1}(x_1) \, h_\gamma \, \Omega (x_2) \; . 
 \label{5.12}
 \ee
Thus for each non intersecting closed path $C$  ---that is,  for each oriented knot $C\subset {\mathbb R}^3$--- with a given  starting and final point $x_0$, the associated holonomy $h_C$ transforms covariantly under gauge transformations, 
\be 
h_C \rightarrow \Omega^{-1}(x_0) \, h_C \, \Omega (x_0) \; . 
\label{5.13}
\ee 
Therefore any function, which is defined on the  $ISU(2)$ conjugacy classes,  determines a classical gauge invariant observable. We shall describe the conjugacy classes of the group $ISU(2)$ in Section~6. For the moment, let us recall the normal construction of classical gauge invariant observables for finite dimensional representations of the structure group. 
Let $[h_C]_\rho$  be the representative of the element $h_C\in ISU(2)$ in the representation $\rho $ of the gauge group. If the  representation $\rho$ is finite dimensional, the 
cyclic property of the trace implies that ${\rm Tr} [h_C]_\rho$  
 is gauge invariant.  Really, in the BF theory we are interested in the $ISU(2)$ representations $(\Lambda, r)$, with $r=0$ or $r=1/2$, which are not finite dimensional. In this case, the ordinary traces of the holonomies in the representation spaces ${\cal H}_{(\Lambda , 0)}$ and ${\cal H}_{(\Lambda , 1/2)}$  need to be improved in order to specify a well defined observable. 
 
\subsection{Trace of holonomies}

  Let us  consider the standard method which is used in physics ---for instance in particle physics and in statistical mechanics--- to describe the sum over the one-particle quantum states. One can introduce appropriately normalized  plane waves 
\be
| \bm k  ) = \frac{1}{\sqrt V} \, e^{i \bm k \bm r} \; , 
\label{5.14}
\ee    
where $V= L^3$ is the volume of a cubic box in which the particle can propagate; then one must consider  the $V \rightarrow \infty $ limit. From the definition (\ref{5.14}) it follows 
\be
( \bm k | \bm k^\prime ) = \frac{(2 \pi )^3}{V} \, \delta^3 (\bm k - \bm k^\prime) \; , 
\label{5.15}
\ee
and
\be
( \bm k | \bm k ) =1 \; . 
\label{5.16}
\ee
With periodic boundary conditions, for instance, the possible values of the momenta are given by $  \bm k = \left ( 2 \pi /L \right ) \bm n  $, with $n_j \in {\mathbb Z}$. Therefore, in the large $L$ limit,  the sum over the eigenstates of the momentum is given by  the integral $  [L^3/ (2 \pi )^3 ]\int d^3 k $, which also coincides with the counting of the number of quantum states in the semiclassical limit by means of the integral $\int d^3p\, d^3q / (2 \pi)^3 $ in classical phase space.  With this notation, the trace of a given operator $O_p $ in the linear space of the one-particle orbital states takes the form   
\be
{\rm Tr } ( O_p ) =  \int {V \, d^3 k \over ( 
2  \pi  )^3}  \, ( \bm k   | \, O_p \,  | \bm k  ) \; ,      
\label{5.17}
\ee
which can easily be controlled in the $V \rightarrow \infty $ limit because of the presence of the overall multiplicative $V$ factor.  

The states of the $(\Lambda , 0 )$ representation are characterized by values of the momentum which belong to the 2-dimensional surface $\bm k^2 = \Lambda^2 $ in momentum space.  In order to make contact with the $\int d^3p \, d^3q / (2 \pi)^3 $ expression for the counting of states in ${\cal H}_{(\Lambda , 0)}$, one can introduce a small thickness $\Delta_P $ to the   $\bm k^2 = \Lambda^2 $ surface. If, for instance, the relation $L \Delta_P / (2 \pi)=1$ is satisfied, then the  $\Delta_P \rightarrow 0 $ limit is recovered in the  $L \rightarrow \infty  $ limit. According to this prescription,  the trace of a given operator $O_p$  in the space ${\cal H}_{(\Lambda , 0)}$ of the   $(\Lambda , 0 )$ representation of $ISU(2)$ reads 
\bea
 {\rm Tr } \left ( O_p \right )  \bigg |_{(\Lambda , 0)} &=&  
 \frac{L^3}{( 2 \pi )^3}\int \left [  d^3 k \right ]_{\bm k^2 \to \Lambda^2 }    \, ( \bm k   | \, O_p \,  | \bm k  ) 
 \nonumber \\ 
&=& \frac{L^2 \Lambda^2  }{(2 \pi )^2} 
   \int d \omega  \, ( \bm k   | \, O_p \,  | \bm k  )
    \quad  , \quad \left (\hbox{ with } \; \ner k \ner k = \Lambda^2 \, \right ) \; ,  
 \label{5.18}
\eea
where $d \omega = \sin \theta \, d \theta \, d \phi $ refers to the solid angle which is defined by the direction of the vector $\bm k $, 
\be
\bm k = \Lambda ( \sin \theta \, \cos \phi ,  \sin \theta \, \sin \phi  , \cos \theta ) \; . 
\label{5.19}
\ee
Note that the presence of the product $L^2 \Lambda^2$  in equation (\ref{5.18}) is required by dimensional reasons. Whereas different prescriptions for the $\Delta_P \rightarrow 0 $ limit may lead to the presence of different adimensional multiplicative factors. These factors play no role because   the Wilson line operators will correspond to appropriately normalized traces.    

In the definition of the  normalized trace of the holonomy $h_C$, the multiplicative factor $L^2 \Lambda^2  /  \pi $  in front of expression (\ref{5.18}) can be removed.   So (in the $L \rightarrow \infty$ limit)  we define the Wilson line operator $W_C$ in the $(\Lambda , 0 )$ representation by means of the normalized trace 
\be
W_C   \, \bigg |_{(\Lambda , 0)} =  \int \frac{d \omega}{4 \pi}  \, ( \bm k   | \, h_C \,  | \bm k  )
    \quad  , \quad \left (\hbox{ with } \; \ner k \ner k = \Lambda^2 \, \right ) \; .  
 \label{5.20}
\ee  
Let us denote the quantum state vectors of a nonrelativistic spin 1/2 particle moving inside a box by $| \ner k )  | s ) = | \ner k ,  s )$,  where $s = \pm 1/2$ refers to the value of one component of the spin.   The normalized trace of the holonomy  $h_C$  in the $(\Lambda , 1/2)$ representation   is defined by 
\be
W_C \, \bigg |_{(\Lambda , 1/2)} =  \sum_s  \int \frac{d \omega}{4 \pi }  \, ( \bm k  , s  | \, h_C \, \pi_+ \,  | \bm k   , s ) 
    \quad  , \quad \left (\hbox{ with } \; \ner k \ner k = \Lambda^2 \, \right )  \; .  
\label{5.21}
\ee
The proof that the BF expectation values of expressions (\ref{5.20}) and (\ref{5.21}) are well defined is reported in Section~7. 

 \section{ $\ner {ISU(2)}$ conjugacy classes}

The set of the conjugacy classes of the inhomogeneous group $ISU(2)$ has rather peculiar properties  that show up also in the values of the corresponding classical characters. 

\subsection{Classes of conjugated elements}
A generic element ${ \cal G }  \in ISU(2) $ can be written as  
\be
{\cal G} = \exp \left [ i \left ( 
\Theta^a J^a + X^a P^a \right ) \right ] = \exp \left [ i \left ( 
\bm \Theta \bm J + \bm X \bm  P \right ) \right ] \; , 
\label{6.1}
\ee
with real parameters $\bm \Theta $ and $\bm X $, in which  $ 0 \leq | \bm \Theta | < 2 \pi $ whereas there are no restrictions on the value of $\bm X$.     Under conjugation with an element of the subgroup $SU(2)$ of $ISU(2)$, the commutation relations  (\ref{2.2}) give 
\be
{ \cal G }  \longrightarrow  e^{- i \beta^a J^a }  \, { \cal G }  \, e^{ i \beta^a J^a } =  \exp \left [ i \left ( 
\bm \Theta^\prime \bm J + \bm X^\prime \bm  P \right ) \right ] \; , 
\label{6.2}
\ee 
where $\bm \Theta^\prime$ and $\bm X^\prime$ denote the rotated vectors 
\be
\left ( \Theta^\prime \right )^a = R^{ab}(\beta) \Theta^b \quad , 
\quad \left ( X^\prime \right )^a = R^{ab}(\beta) X^b \; , 
\label{6.3}
\ee
which are obtained according to the  adjoint representation of $SU(2)$, {\it i.e.} $R^{ab}(\beta) \in SO(3)$.   Therefore, the conjugacy class of  ${ \cal G } $ is possibly labelled by the rotation invariants  $| \bm \Theta |$, $| \bm X |$ and $ \bm \Theta \bm X = \Theta^a X^a$. On the other hand, under conjugation with a translation element of $ISU(2)$ 
\be
{ \cal G }   \longrightarrow  e^{- i \eta^a P^a }  \, { \cal G }  \, e^{ i \eta^a P^a } =  \exp \left [ i \left ( 
\widetilde {\bm \Theta} \bm J + {\widetilde {\bm X}} \bm  P \right ) \right ] \; , 
\label{6.4}
\ee
one finds 
\be
\widetilde \Theta^a =   \Theta^a \quad , \quad  \widetilde X^a =  X^a + \epsilon^{abc} \eta^b \Theta^c \; . 
\label{6.5} 
\ee
Equation (\ref{6.5}) shows that the parameter $\bm \Theta $ is not modified and 
\begin{itemize}
\item when $\bm \Theta =0 $,  $\bm X $ is not modified; 
\item   when  $\bm \Theta \not=0 $,  the component of $\bm X$ which is orthogonal to $\bm \Theta $ can be arbitrarily modified. While the component of $\bm X$ along the direction of $\bm \Theta$ is not modified. 
\end{itemize}
Consequently,  equations (\ref{6.3}) and (\ref{6.5}) show that the conjugacy classes of $ISU(2)$ can be labelled by two real numbers $(r_1 , r_2 )$ with  $ r_1 = | \bm \Theta | $ and 

\begin{itemize} 
\item  $r_2 = | \bm X | $, when $r_1 = 0$; 
\item $ r_2 = \bm \Theta \bm X $, when $r_1 \not= 0$. 
\end{itemize}

\noindent The set of variables $\{ (r_1, r_2) \}$ does not parametrize a two dimensional manifold because of the singularity  at $r_1 = 0$.  

\subsection{Classical traces}

Let $ {\rm Tr } \left ( { \cal G }  \right )  \big |_{(\Lambda , r)}$ be  the trace of ${ \cal G }  \in ISU(2)$ in the $(\Lambda , r)$ representation of $ISU(2)$ (with $r=0 , 1/2$), 
\be
 {\rm Tr } \left ( { \cal G }  \right )  \bigg |_{(\Lambda , r)}  = 
  \left  \{ \begin{array}  {l@{ ~ } l}  
   \frac{L^3}{( 2 \pi )^3}\int \left [  d^3 k \right ]_{\bm k^2 \to \Lambda^2 }     \, ( \bm k   | \, { \cal G }  \,  | \bm k  ) &  \quad \hbox{ when } \; r=0 \; ;    \\   ~ & ~ \\  
     \frac{L^3}{( 2 \pi )^3} \sum_s \int \left [  d^3 k \right ]_{\bm k^2 \to \Lambda^2 }    \, ( \bm k  , s  | \, { \cal G }  \, \pi_+ \,  | \bm k   , s )
 &  \quad \hbox{ when }  \;  r=1/2 \; .   
\end{array} \right. 
\label{6.6}
\ee 
   By means of equations   (\ref{5.15}), (\ref{5.16}) and (\ref{5.18}) one finds 
 \begin{description}
\item[{(1)} ] When $ \bm \Theta = 0$ and $\bm X = 0 $, 
\be
 {\rm Tr } \left ( { \cal G }  \right )  \bigg |_{(\Lambda , r)} = \frac{L^2 }{ ( 2\pi )^2} \, 4 \pi \Lambda^2  \;  . 
 \label {6.7}
\ee

\item[{(2)} ]  When $ \bm \Theta = 0$ and $\bm X \not= 0 $, 
\be
 {\rm Tr } \left ( { \cal G }  \right )  \bigg |_{(\Lambda , r)} = \frac{L^2 }{ ( 2\pi )^2} \, 4 \pi \Lambda^2  \,  \frac{\sin (\Lambda |\bm X|)}{\Lambda |\bm X|}\;  . 
 \label {6.8}
\ee
 
 \item[{(3)} ]  When $ \bm \Theta \not= 0$ and $\bm X = 0 $,  let $ | \bm k^\prime  )= e^{i \Theta^a J^a} \,  | \bm k    ) $. 
One has   
 \be
 ( \bm k  | \, e^{i \Theta^a J^a} \,  | \bm k    ) = \frac{( 2 \pi )^3}{V } \, \delta \left ( \bm k  - \bm k^\prime  \right )  \; .  
 \label{6.9}
 \ee
Since $( \bm k  |   \bm k^\prime  ) =  ( \bm k  | \, e^{i \Theta^a J^a} \, | \bm k ) $ is vanishing unless the vector $ \bm k$ is directed as $\pm \bm \Theta $, with $\bm \Theta = (\Theta^1, \Theta^2 , \Theta^3)$,   one obtains 
 \be
 {\rm Tr } \left ( { \cal G }  \right )  \bigg |_{(\Lambda , 0)} = 2 \;  , 
 \label {6.10}
\ee
which is in agreement with the Frobenius fixed point theorem \cite{SS} since any nontrivial rotation of a spherical surface in ${\mathbb R}^3$  has just two fixed points. In the case of the $(\Lambda , 1/2)$ representation, one finds 
\be
 {\rm Tr } \left ( { \cal G }  \right )  \bigg |_{(\Lambda , 1/2)} = 2 \, \cos \left ( | \bm \Theta | / 2 \right ) \;  .  
 \label {6.11}
\ee
 
\item[{(4)} ]  When $ \bm \Theta \not= 0$ and $\bm X \not= 0 $, 
 \be
 {\rm Tr } \left ( { \cal G }  \right )  \bigg |_{(\Lambda , r)} = 2 \, \cos \left (  \Lambda \bm X  \bm {\widehat  \Theta } +   r| \bm \Theta |  \right ) \;  ,  
 \label {6.12}
\ee
 where the unit vector $\ner {\widehat  \Theta} $ is defined by 
$  \ner {\widehat  \Theta} = \ner \Theta / |\ner \Theta |   $.
 
\end{description}

\noindent The observed discontinuity of the classical trace of ${ \cal G } $ at $\Theta =0 $ matches   the structure of the set of $ISU(2)$ conjugacy classes discussed in Section~6.1.  
 
\section{Expectation values}

Let us concentrate on the BF topological invariants which are associated with oriented framed coloured knots in ${\mathbb R}^3$.  A  knot $C$ in ${\mathbb R}^3$, with a specified   irreducible $ISU(2)$ representation,   is called a coloured knot.    The invariant $\langle W_C \rangle$ which is associated with the  knot $C$ is defined by the BF expectation value of the Wilson line operator  
\be
\langle W_C \rangle   = {\int D\hbox{(fields) } e^{iS_{TOT}} \, W_{C }  \over 
\int D\hbox{(fields) } e^{iS_{TOT}}  }  \; ,  
\label{7.1}
\ee
where $W_C$ corresponds to the normalized trace of the holonomy $h_C$ shown in equations   (\ref{5.20}) and (\ref{5.21}).  In perturbation theory, the determination of $\langle W_C \rangle$ is obtained by means of the following steps: (1) expansion of the holonomy $h_C$ in powers of the gauge fields, (2) computation  of the vacuum expectation values of the products of the gauge fields, and  (3)  evaluation of  the normalized trace of the $ISU(2)$ generators.  

In the quantum BF field theory,  the holonomy $h_C$ is a composite operator and its expansion in powers of the connection $\cal A$ contains product of fields at coincident points. As in the case of the quantum Chern-Simons field theory, the ambiguities  of the mean value (\ref{7.1}),  which are due to the presence of fields at coincident points,  are removed by means of the point-splitting limit procedure \cite{GMM,E} which is based on the introduction of a framing of the knot $C$.   So, the invariant (\ref{7.1}) is really defined for framed  knots. 

The perturbative computation of $\langle W_C \rangle$ is based on the expansion of $h_C $ in powers of the fields
\bea
h_C &=& 1 + i \int_C  {\cal A}_\mu(x) dx^\mu + i^2 \int_C dx^\mu \int^x_{x_0} dy^\nu \, {\cal A}_\nu(y) {\cal A}_\mu(x) \nonumber \\ 
&& + i^3 \int_C dx^\mu \int^x_{x_0} dy^\nu \int^y_{x_0} dz^\lambda \, {\cal A} _\lambda (z) {\cal A}_\nu(y) {\cal A}_\mu(x) + \cdots 
\label{7.2}
\eea
where ${\cal A}_\mu(x) = A^a_\mu (x) J^a + B^a_\mu (x) P^a$ and $x_0$ denotes a  given base point on the oriented knot $C$.  In expression (\ref{7.2}), it is understood that the generators $\{ J^a , P^b \}$ are multiplied according to  the order shown in the formula. More precisely, if $\{ J^a , P^b \}$ are collectively denoted by $\{ T^\alpha \}$, one has   ${\cal A}_\mu(x) = {\cal A}^\alpha_\mu  (x) T^\alpha $ and in equation (\ref{7.2}) the products of connections mean, for instance,  
$$ 
\left [ {\cal A}_\nu(y) {\cal A}_\mu(x)\right ]_{ij}  =  {\cal A}^\beta_\nu(y) {\cal A}^\alpha_\mu(x) \, T^\beta_{ik}  T^\alpha_{kj} \; , 
 $$
$$ 
\left [ {\cal A} _\lambda (z) {\cal A}_\nu(y) {\cal A}_\mu(x)\right ]_{ij}  =  {\cal A} _\lambda^\gamma (z) {\cal A}^\beta_\nu(y) {\cal A}^\alpha_\mu(x) \, T^\gamma_{i\ell }T^\beta_{ \ell k}  T^\alpha_{kj} \ . 
 $$
 When the $ISU(2) $ generators are not multiplied, they can be understood as elements of a tensor product in colour space; so,   it is convenient to introduce the notation 
 \bea
 {\cal A}_\mu (x) \otimes {\cal A}_\nu(y) &=& {\cal A}^\alpha_\mu (x) {\cal A}^\beta_\nu (y)\, T^\alpha_{ij} T^\beta_{k \ell } \nonumber \\
  {\cal A}_\mu (x) \otimes {\cal A}_\nu (y) \otimes {\cal A}_\lambda (z) &=& {\cal A}_\mu^\alpha (x) {\cal A}_\nu^\beta (y) {\cal A}_\lambda^\gamma (z) \, T^\alpha_{ij} T^\beta_{k \ell } T^\gamma_{mn}
 \; \; , ... \hbox{ etc.}  
 \label{7.3}
 \eea
 
 According to equation (\ref{7.2}), for each $ISU(2)$ representation $(\Lambda , r) $ with $r=0$ or $r=1/2$,   the normalized  trace of $h_C$ in the colour space takes the form of a sum of normalized traces of product of generators $J^a$ and $P^b$. It should be noted that, since the representations $(\Lambda , r)$ are infinite dimensional,   the cyclic property of the trace   is no more valid; consequently,  the classical gauge invariance of the trace of $h_C$ is not guaranteed. What saves the day   is  that  the field theory expectation values  of  connection's products are invariant under global $ISU(2)$ transformations.  

\begin{prop}
The {\rm BF} expectation values computed by means of  the total action $S_{TOT} = S + S_{\phi \pi}$ satisfy 
\bea
\langle {\cal A}_\mu(x_1) \otimes {\cal A}_\nu(x_2) \otimes \cdots \otimes {\cal A}_\lambda (x_n)\rangle  &&=    \nonumber \\
 &&{\hskip - 5 cm}= 
\langle \left ( {\cal G}^{-1}{\cal A}_\mu (x_1) {\cal G} \right ) \otimes \left ( {\cal G}^{-1} {\cal A}_\nu (x_2) {\cal G}\right ) \otimes \cdots \otimes \left ( {\cal G}^{-1}{\cal A}_\lambda (x_n) {\cal G} \right ) \rangle  \; , 
\label{7.4}
\eea
 for any ${\cal G} \in ISU(2)$. 
 
 \end{prop}

\noindent {\bf Proof.}
The proof is made of  two parts. First it shown that equation (\ref{7.4}) is satisfied in the case in which  ${\cal G} = e^{i \beta^a J^a}$, and then it is demonstrated that equality (\ref{7.4}) is satisfied for ${\cal G} = e^{i \eta^a P^a}$. 

When ${\cal G} = e^{i \beta^a J^a}$, one has 
\be
 {\cal G}^{-1}{\cal A}_\mu (x) {\cal G} = A^{ \prime a}_\mu (x) J^a + B^{\prime a}_\mu (x) P^a\; , 
 \label{7.5}
\ee
where 
\be
A^{ \prime a}_\mu (x)  = R^{ab}(\beta) A^b_\mu (x) \quad , 
\quad B^{ \prime a}_\mu (x) = R^{ab} (\beta ) B^b_\mu (x)  \; ,  
\label{7.6}
\ee
with $R^{ab}(\beta) \in SO(3)$. Under the change of variables $A^a_\mu (x) \rightarrow A^{ \prime a}_\mu (x) $, $B^a_\mu (x) \rightarrow B^{ \prime a}_\mu (x)$ and 
\bea
M^a(x) \rightarrow  R^{ab}(\beta)M^b(x)  \quad & , &  \quad N^a(x) \rightarrow  R^{ab}(\beta)N^b(x) \; , \nonumber \\ 
c^a(x) \rightarrow  R^{ab}(\beta)c^b(x) \quad  & , &  \quad {\overline  c}^{\, a}(x) \rightarrow  R^{ab}(\beta){\overline c}^{ \, b}(x) \; ,  \nonumber \\ 
\xi^a(x) \rightarrow R^{ab}(\beta)\xi^b(x) \quad & , &  \quad {\overline  \xi}^{\, a}(x) \rightarrow  R^{ab}(\beta){\overline \xi }^{ \, b}(x) \;  ,  
\label{7.7}
\eea
the total action $S_{TOT} = S + S_{\phi \pi}$ is invariant.  
Therefore equation  (\ref{7.4}) is fulfilled when  ${\cal G} = e^{i \beta^a J^a}$. 

In the case ${\cal G} = e^{i \eta^a P^a}$, one gets  
\be
 {\cal G}^{-1}{\cal A}_\mu (x) {\cal G} = {\widetilde A}^a_\mu (x) J^a + {\widetilde B}^{a}_\mu (x) P^a\; , 
 \label{7.8}
\ee
where 
\be
{\widetilde A}^{a}_\mu (x)  =  A^a_\mu (x) \quad , 
\quad {\widetilde B}^{a}_\mu (x) =   B^a_\mu (x) + \epsilon^{abc} \eta^b A_\mu^a (x)\; .   
\label{7.9}
\ee
Under the change of variables $A_\mu^a(x) \rightarrow {\widetilde A}^{a}_\mu (x)$, $B^a_\mu (x) \rightarrow {\widetilde B}^{a}_\mu (x) $ and 
\bea
M^a(x) \rightarrow  M^a(x) - \epsilon^{abc } N^b \eta^c  \quad & , &  \quad N^a(x) \rightarrow  N^a(x) \; , \nonumber \\ 
c^a(x) \rightarrow  c^a(x) \quad  & , &  \quad {\overline  c}^{\, a}(x) \rightarrow  {\overline c}^{ \, a}(x) - \epsilon^{abc} {\overline \xi}^b \eta^c \; ,  \nonumber \\ 
\xi^a(x) \rightarrow \xi^a(x) - \epsilon^{abc} c^b \eta^c \quad & , &  \quad {\overline  \xi}^{\, a}(x) \rightarrow  {\overline \xi }^{ \, a}(x) \;  ,  
\label{7.10}
\eea
the total action $S_{TOT} = S + S_{\phi \pi}$ is  invariant as a consequence of the Jacobi identity. Thus equation (\ref{7.4}) is satisfied for ${\cal G} = e^{i \eta^a P^a}$. 

To sum up, equation (\ref{7.4}) is satisfied when ${\cal G} = e^{i \beta^a J^a}$ with arbitrary $\beta^a$ and also when ${\cal G} = e^{i \eta^a P^a}$ with arbitrary $\eta^a$. Therefore equality  (\ref{7.4}) holds  for any ${\cal G} \in ISU(2) $. {\hfill \cvd} 

\medskip 

A first consequence of equation (\ref{7.4}) is that the two-points function $\langle A^a_\mu (x) A^b_\nu (y) \rangle $ must vanish because there is not an $ISU(2)$ invariant which is quadratic in $J^a$. 

In the expansion (\ref{7.2}) of $h_C$ in powers of the fields, the generators of $ISU(2)$ are multiplied; hence equation (\ref{7.4}) implies 
\be
\langle h_C \rangle = {\cal G}^{-1}  \langle h_C \rangle {\cal G} \; , \qquad \forall {\cal G} \in ISU(2) \; .  
\label{7.11}
\ee
Thus, as in the case of the Chern-Simons theory,  the expectation value of the holonomy associated with a knot $C$ ---with colour given by an irreducible representation of the gauge group--- is proportional to the identity in colour space or, more precisely, it is a function of the Casimir operators of the structure group.  This means that $\langle W_C \rangle$,  which is the normalized trace of $\langle h_C \rangle $ in the $ISU(2)$ representations $(\Lambda , 0)$ and $(\Lambda , 1/2)$, is well defined, it is gauge invariant and it does not depend on the choice of the base point on $C$. 

Finally,   since  the holonomy $h_C$  does not depend on the metric of ${\mathbb R}^3$ and the only dependence of the total action on the metric  is contained in the gauge fixing terms, the expectation value  (\ref{7.1}) corresponds to a topological invariant of oriented framed coloured knots  in ${\mathbb R}^3$.    

\section{Perturbative expansion of the observables}

The value of the observable $\langle W_{C} \rangle$, which  is given by the normalized trace of the expectation  value of the holonomy associated with the knot $C \subset {\mathbb R}^3$, 
\be
\langle W_C   \rangle \, \bigg |_{(\Lambda , 0)} =  \int \frac{d \omega}{4 \pi}  \, ( \bm k   | \, \langle h_C \rangle \,  | \bm k  )
    \quad  , \quad \left (\hbox{ with } \; \ner k \ner k = \Lambda^2 \, \right ) \; .  
 \label{8.1}
\ee  
\be
\langle W_C \rangle \, \bigg |_{(\Lambda , 1/2)} =  \sum_s  \int \frac{d \omega}{4 \pi }  \, ( \bm k  , s  | \, \langle h_C \rangle \, \pi_+ \,  | \bm k   , s ) 
    \quad  , \quad \left (\hbox{ with } \; \ner k \ner k = \Lambda^2 \, \right )  \; .  
\label{8.2}
\ee
can be  obtained by computing the expectation value  $ \langle h_C \rangle$ by means of an expansion of  $  h_C $ in powers of the fields. It is important to note that, in the evaluation of $ \langle h_C \rangle$, the presence of a base point $x_0$ in the knot $C$ must be taken into account. Thus, $\langle W_{C} \rangle$ takes the form of a sum of an infinite number of perturbative contributions. 

The invariant  $\langle W_C \rangle$ can be approximated by considering only a finite number of terms, but the  truncation of the perturbative series cannot be  introduced   arbitrarily. In order to obtain topological invariants, one needs to sum all the diagrams which are necessary to ensure the gauge invariance of the result. This can be achieved by summing all the diagrams which are of the same order in powers of the  parameters which multiply the gauge invariant lagrangian terms.    The BF action is the sum of two terms which are separately invariant under gauge transformations; so,  two independent parameters are required. 
One parameter can be chosen to be $\hbar $,  and the second parameter can be taken to be  the coupling constant $g$. 

In the previous sections, the  convention  $\hbar =1$ has been used. In this section,  the dependence of the Feynman diagrams on $\hbar$ is made explicit.   Let us recall that  a given Feynman diagram with $\cal P$ propagators and $\cal V$ interaction vertices is of order $\hbar^{{\cal P}-{\cal V}}$. The dependence of a diagram on the coupling constant $g$   can easily be determined because  $g$ multiplies the $B B$ component of the propagator, equation (\ref{2.12}), and the $AAA$ interaction lagrangian term.   

Note that  the possible values of the group generators $J^a$ and $P^b$ represent  ``colour quantum numbers'' that  have  vanishing  field theory dimensions. If one wishes to give a physical interpretation to the vectors of  the $ISU(2)$ representations as particle state vectors, one can imagine that the eigenvalues of ``momentum" $P^a $ refer to a given momentum scale, so that $\Lambda$ is dimensionless. 

In what follows, the perturbative  contributions to $\langle W_{C} \rangle$ of order $\hbar^n $ with $n=0,1,2,3$  are in order. The contribution  of  order  $\hbar^n$ is indicated by  $\langle W_{C} \rangle^{(n)}$ and contains all the nonvanishing components which are labelled by  powers of $g$.  The colour of the knot is specified by the $(\Lambda , r)$ representation of $ISU(2)$ with $r=0,1/2$. 

\subsection{Lowest order}
With the chosen normalization of the traces shown in equations (\ref{8.1}) and (\ref{8.2}), the component  of $\langle W_C \rangle$ of order $\hbar^0$ is just the unit 
\be
\langle W_C \rangle^{(0)}  = 1   \; . 
\label{8.3}
\ee
 
\subsection{First order} 

The contributions of order $\hbar$ are given  by the integration of the two components  of the  field propagator along the knot  $C$, as sketched in Figure~6. The double line of Figure~6 generically indicates a framed knot $C$ with its base point $x_0$ pointed out. The  embedding of $C$ in ${\mathbb R}^3$ is not shown.  A simple line represents a gauge field propagator \eqref{2.12}. 

\vskip 0.6 truecm

\centerline {
\begin{tikzpicture} [scale=0.9] [>=latex]
\draw [very thick  ] (0,0) circle (1.4); 
\draw [very thick  ] (0,0) circle (1.5); 
\draw [very thick]  (-1.4,0) -- (1.4,0); 
\draw [thick] (0,-1.2) -- (0,-1.7);
\node at (0.1,-2.1) {$x_0$};
\node at (1.6,1.3) {$C$};
\end{tikzpicture}
}

\vskip 0.3 truecm
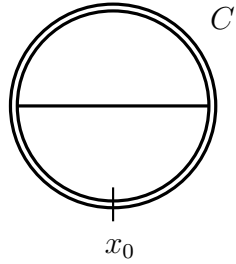
\centerline {{Figure 6.} {First order contribution to $\langle W_C \rangle$.}}

\vskip 0.4 truecm

\noindent In this case, the point-splitting procedure, which is defined by means of the framing $C_f$ of the knot $C$, is used.  Since the $AB$ component of the propagator is of order $\hbar$ and the $BB$ component of the propagator is of order $\hbar g$,   one finds
\be
\langle W_C \rangle^{(1)}  =  - i \left ( {\hbar \over 2} \right ) \, \ell k (C,C_f) \left ( 2 \Lambda r - g \Lambda^2 \right ) \; , 
\label{8.4}
\ee
where $\ell k (C,C_f)$ denotes the linking number of $C$ and its framing $C_f$. Indeed, the linking number of two oriented knots $C_1$ and $C_2$  can be expressed \cite{R} by means of  the Gauss integral   
\be
\ell k (C_1,C_2) = \frac{1}{4 \pi} \oint_{C_1} dx^\nu \oint_{C_2} dy^\sigma \epsilon_{\nu \sigma \lambda } \frac{(x-y)^\lambda}{|x - y|^3} \; . 
\label{8.5}
\ee

\subsection{Second  order} 

 The nonvanishing contributions of order $\hbar^2$  to $\langle W_C \rangle$ are related with diagrams with two field propagators, shown in Figure~7, and diagrams with one vertex and three  field propagators shown in Figure~8. As shown in Section~4, diagrams with one loop give vanishing results of order  $\hbar^2$.   
 
 In the computation of $\langle h_C \rangle$,  diagrams with two field propagators give contributions which are proportional to the  combinations of Casimir operators: $(JP)^2= (J^a P^a)^2$, $(JP) (P^2) =(J^a P^a)(P^bP^b)$ and $ (P^2)^2 = (P^a P^a)^2$. Moreover,   from the diagrams of the type shown in the second picture of Figure~7, one gets an additional  contribution which is proportional to the Casimir operator $P^a P^a$. This is a consequence of the identity  
 \be
 P^a J^b J^a P^b = (J^a P^a)^2 - 2 (P^a P^a) \; , 
 \label{8.6}
 \ee
 which follows from the structure of the  $ISU(2)$ algebra. 
 
  \vskip 0.6 truecm

\centerline {
\begin{tikzpicture} [scale=0.9] [>=latex]
\draw [very thick  ] (0,0) circle (1.4); 
\draw [very thick  ] (0,0) circle (1.5); 
\draw [very thick]  (-1.3,0.5) -- (1.3,0.5); 
\draw [very thick]  (-1.3,-0.5) -- (1.3,-0.5); 
\draw [thick] (0,-1.2) -- (0,-1.7);
\node at (0.1,-2.1) {$x_0$};
%
%
\begin{scope}[shift={(5,0)}]
\draw [very thick  ] (0,0) circle (1.4); 
\draw [very thick  ] (0,0) circle (1.5); 
\draw [very thick]  (-1.28,0.6) -- (1.28,-0.6); 
\fill[white] (-0.25,-0.19) rectangle (0.25, 0.19);
\draw [very thick]  (-1.28,-0.6) -- (1.28,0.6);  
\draw [thick] (0,-1.2) -- (0,-1.7);
\node at (0.1,-2.1) {$x_0$};
\end{scope}
\begin{scope}[shift={(10,0)}]
\draw [very thick  ] (0,0) circle (1.4); 
\draw [very thick  ] (0,0) circle (1.5); 
\draw [very thick]  (0.5,-1.3) -- (0.5,1.3); 
\draw [very thick]  (-0.5,-1.3) -- (-0.5,1.3); 
\draw [thick] (0,-1.2) -- (0,-1.7);
\node at (0.1,-2.1) {$x_0$};
\end{scope}
\end{tikzpicture}
}

\vskip 0.3 truecm
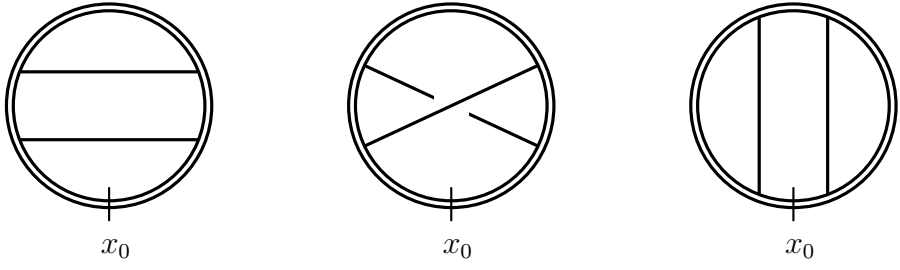
\centerline {{Figure 7.} {Second order contribution to $\langle W_C \rangle$ with two field propagators.}}

\vskip 0.4 truecm

The contributions to   $\langle h_C \rangle$ coming from the diagrams of Figure~7 are 
\bea
 &&- \frac{\hbar^2}{2} \left (  \oint_{C} dx^\nu \oint_{C_f} dy^\sigma \epsilon_{\nu \sigma \lambda } \frac{(x-y)^\lambda}{4 \pi |x - y|^3} \right )^2 \left [ (JP)^2  - g (JP) P^2 + \frac{ g^2}{4} (P^2 )^2\right ] + \nonumber \\ 
 && ~~ + 2 P^2 \hbar^2 \oint_C dx^\mu \int^x_{x_0} dy^\nu \int^y_{x_0} dz^\lambda \int^z_{x_0} dw^\sigma  \, \frac{\epsilon_{\nu \sigma \tau} \epsilon_{\lambda \mu \rho } (y-w)^\tau (z-x)^\rho   }{16 \pi^2 |y-w|^3 |z-x|^3 } \; . 
 \label{8.7}
\eea

 \vskip 0.6 truecm

\centerline {
\begin{tikzpicture} [scale=0.9] [>=latex]
\draw [very thick  ] (0,0) circle (1.4); 
\draw [very thick  ] (0,0) circle (1.5); 
\draw [very thick]  (-1.2,-0.7) -- (0,0); 
\draw [very thick]  (1.2,-0.7) -- (0,0);
\draw [very thick] (0,0) -- (0,1.4);
\draw [thick] (0,-1.2) -- (0,-1.7);
\node at (0.1,-2.1) {$x_0$};
%
%
%
%
\end{tikzpicture}
}

\vskip 0.3 truecm
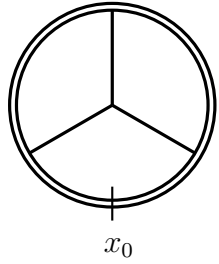
\centerline {{Figure 8.} {Second order contribution to $\langle W_C \rangle$ with one vertex.}}

\vskip 0.4 truecm

 \noindent The nonvanishing contribution to $\langle h_C \rangle$ coming from the diagram of Figure~8 is proportional to the Casimir operator $P^2$, as a consequence of the identity 
 \be
 \epsilon^{abc} \, P^b J^a P^c = - 2 i P^a P^a \; ,  
 \label{8.8}
 \ee
 and takes the form 
 \be
 2 P^2 \hbar^2 \int d^3x  \oint_C dz^\sigma \int^z_{x_0} du^\tau \int^u_{x_0} dv^\rho  \, \epsilon^{\mu \nu \lambda } \epsilon_{\mu \tau \xi} \epsilon_{\nu \rho \beta} \epsilon_{\lambda \sigma \alpha }
 \frac{(x-u)^\xi (x-v)^\beta (x-z)^\alpha}{64 \pi^3 |x-u|^3 |x-v|^3 |x-z|^3}\, . 
 \label{8.9}
 \ee
 The sum of all the terms of order $\hbar^2$ is given by 
\be
\langle W_C \rangle^{(2)}  =  - \frac{1}{2} \left ( {\hbar \over 2} \right )^2 \, \left [\, \ell k (C,C_f) \, \right ]^2 \left ( 2 \Lambda r - g \Lambda^2 \right )^2 + \hbar^2 \Lambda^2 \rho(C)\; ,   
\label{8.10}
\ee
where $\rho (C)$ is the knot invariant that has been found \cite{GMM} in the study of the knot polynomials which are derived from the Chern-Simons field theory,   
\bea
\rho (C) &=& \oint_C dx^\mu \int^x_{x_0} dy^\nu \int^y_{x_0} dz^\lambda \int^z_{x_0} dw^\sigma  \, \frac{\epsilon_{\nu \sigma \tau} \epsilon_{\lambda \mu \rho } (y-w)^\tau (z-x)^\rho   }{8 \pi^2 |y-w|^3 |z-x|^3 } \nonumber \\ 
&& +    \oint_C dz^\sigma \int^z_{x_0} du^\tau \int^u_{x_0} dv^\rho  \, \epsilon^{\mu \nu \lambda } \epsilon_{\mu \tau \xi} \epsilon_{\nu \rho \beta} \epsilon_{\lambda \sigma \alpha }\, 
\der^\xi_u \, \der^\beta_v \, {\cal I}^\alpha \; , 
\label{8.11}
 \eea
where
\be
{\cal I}^\alpha = \frac {  | v-u| + | z- u| - |v-z|}{ 16 \pi^2 \left (  |v-u|\, |z-u| + (v-u)(z-u) \right )}  \left [ \frac{(v-u)^\alpha}{|v-u|} + \frac{(z-u)^\alpha}{|z-u|} \right ] \; . 
\label{8.12}
\ee
The $\rho (C)$ knot invariant \cite{GMM} gives the analytic expression  of the second coefficient of the Alexander-Conway polynomial  \cite{R,AL,CON}. 

\subsection{Third order}

The value of $\langle W_C \rangle^{(3)}$ is given by the sum of the amplitudes which are associated with diagrams containing  $3$, $4$ and $5$ field propagators (\ref{2.12}). 
In the computation of $\langle h_C \rangle$ at order $\hbar^3$, diagrams with one loop produce vanishing results. 
The contributions corresponding to the diagrams with  $5$ propagators and two lagrangian vertices, shown in Figure~9, are vanishing as a consequence of the   algebra structure (\ref{2.2}) of the $ISU(2)$ generators. 

 \vskip 0.6 truecm

\centerline {
\begin{tikzpicture} [scale=0.9] [>=latex]
\draw [very thick  ] (0,0) circle (1.4); 
\draw [very thick  ] (0,0) circle (1.5); 
\draw [very thick]  (-1,-1) -- (0,-0.5); 
\draw [very thick]  (1,-1) -- (0,-0.5);
\draw [very thick]  (-1,1) -- (0,0.5); 
\draw [very thick]  (1,1) -- (0,0.5);
\draw [very thick] (0,-0.5) -- (0,0.5);
%
%
\node at (4,0) {$+$ ~permutations};
\draw [thick] (0,-1.2) -- (0,-1.7);
\node at (0.1,-2.1) {$x_0$};
\end{tikzpicture}
}

\vskip 0.3 truecm
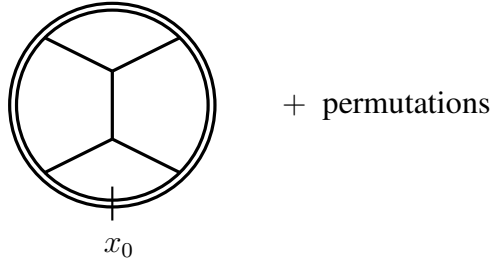
\centerline {{Figure 9.} {Third order diagrams with two  vertices.}}

\vskip 0.4 truecm

\noindent Diagrams with $4$ propagators contain one vertex and are  of the type shown in Figure~10. The corresponding amplitudes contain the combinations $(JP) P^2$ and $(P^2)^2$ of the Casimir operators. The sum of these contributions to $\langle h_C \rangle $ is given by 
\bea
&& P^2 \hbar^3  \int d^3w  \oint_C dz^\sigma \int^z_{x_0} du^\tau \int^u_{x_0} dv^\rho  \, \epsilon^{\mu \nu \lambda } \epsilon_{\mu \tau \xi} \epsilon_{\nu \rho \beta} \epsilon_{\lambda \sigma \alpha }
 \frac{(w-u)^\xi (w-v)^\beta (w-z)^\alpha}{ |w-u|^3 |w-v|^3 |w-z|^3}
\times \nonumber \\ 
&& \times \frac{-i}{128 \pi^4}  \left [ (JP) - \frac{g}{2}  P^2  \right] \oint_C dx^\mu \oint_{C_f} dy^\nu \epsilon_{\mu \nu 
\lambda} \frac{(x-y)^\lambda }{|x-y|^3 } \, . 
\label{8.13}
\eea

 \vskip 0.6 truecm

\centerline {
\begin{tikzpicture} [scale=0.9] [>=latex]
\draw [very thick  ] (0,0) circle (1.4); 
\draw [very thick  ] (0,0) circle (1.5); 
\draw [very thick]  (-1.28,-0.6) -- (1.28,-0.6); 
\draw [very thick]  (-1.4,0.1) -- (1.4,0.1);
\draw [very thick]  (0,0.1) -- (0,1.4); 
%
%
\node at (4,0) {$+$ ~permutations};
\draw [thick] (0,-1.2) -- (0,-1.7);
\node at (0.1,-2.1) {$x_0$};
\end{tikzpicture}
}

\vskip 0.3 truecm
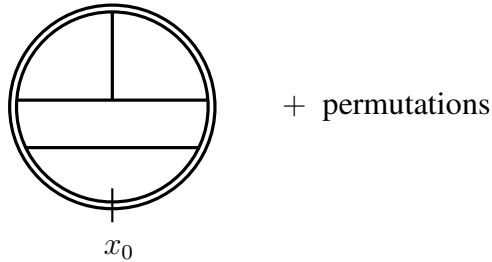
\centerline {{Figure 10.} {Third order diagrams with four propagators.}}

\vskip 0.4 truecm

\noindent Diagrams with $3$ propagators are sketched in Figure~11. The combinations of Casimir operators that one finds in this case are $(JP)^3$, $(JP)^2 P^2$, $(JP) (P^2)^2$, $(P^2)^3$, $(JP) P^2 $ and $(P^2)^2$. The resulting  $\langle h_C \rangle $ amplitude which is associated with the diagrams of Figure~11 is given by 
\bea
&&\frac{i \hbar^3}{6} \left [ (JP) - \frac{g}{2} P^2 \right ]^3 \left (  \oint_{C} dx^\nu \oint_{C_f} dy^\sigma \epsilon_{\nu \sigma \lambda } \frac{(x-y)^\lambda}{4 \pi |x - y|^3} \right )^3 + \nonumber \\ 
&&+ \int d^3x  \oint_C dz^\sigma \int^z_{x_0} du^\tau \int^u_{x_0} dv^\rho  \, \epsilon^{\mu \nu \lambda } \epsilon_{\mu \tau \xi} \epsilon_{\nu \rho \beta} \epsilon_{\lambda \sigma \alpha }
 \frac{(x-u)^\xi (x-v)^\beta (x-z)^\alpha}{64 \pi^3 |x-u|^3 |x-v|^3 |x-z|^3} \times \nonumber \\ 
&&\times \left ( -i 2\hbar^3 \right ) \left [ (JP) - \frac{g}{2} P^2 \right ] P^2 \left (  \oint_{C} dx^\nu \oint_{C_f} dy^\sigma \epsilon_{\nu \sigma \lambda } \frac{(x-y)^\lambda}{4 \pi |x - y|^3} \right )\, . 
\label{8.14}
\eea

 \vskip 0.6 truecm

\centerline {
\begin{tikzpicture} [scale=0.9] [>=latex]
\draw [very thick  ] (0,0) circle (1.4); 
\draw [very thick  ] (0,0) circle (1.5); 
\draw [very thick]  (-0.81,1.15) -- (1.04,-0.92); 
\fill[white] (-0.06,-0.01) rectangle (0.28, 0.3);
\fill[white] (0.4,-0.6) rectangle (0.7, -0.25);
\draw [very thick]  (-1.35,-0.4) -- (1.35,-0.4); 
\fill[white] (-0.71,-0.6) rectangle (-0.24, -0.25);
\draw [very thick]  (-1.01,-0.95) -- (0.96,1.01);  
\node at (4,0) {$+$ ~permutations};
\draw [thick] (0,-1.2) -- (0,-1.7);
\node at (0.1,-2.1) {$x_0$};
\end{tikzpicture}
}

\vskip 0.3 truecm
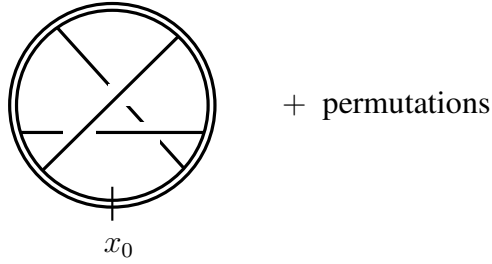
\centerline {{Figure 11.} {Third order diagrams with three  propagators.}}

\vskip 0.4 truecm

\noindent Finally, the sum of all the contributions of order $\hbar^3$ takes the form 
\bea
\langle W_C \rangle^{(3)}  &=&   \frac{i}{6} \left ( {\hbar \over 2} \right )^3  \left ( 2 \Lambda r - g \Lambda^2 \right )^3 \, \left [\, \ell k (C,C_f) \, \right ]^3 + \nonumber \\ 
&& \quad -i \frac{\hbar^3}{2}   \left ( 2 \Lambda r - g \Lambda^2 \right )  \Lambda^2 \, \left [\, \ell k (C,C_f) \, \right ] \, \rho(C)\; .    
\label{8.15}
\eea

\subsection{Chern-Simons comparison}

The knot invariants contained in $\langle W_C \rangle^{(1)} $ and $\langle W_C \rangle^{(2)} $ are precisely the invariants that one also finds  in the Chern-Simons field theory (multiplying different Casimir operators, of course). At the third order, the knot invariants of the BF and of the Chern-Simos theory differ significantly.  Indeed, the third order  term $\langle W_C \rangle^{(3)} $ in the Chern-Simons theory  ---which has been computed correctly by Hirshfeld and Sassenberg  \cite{HS}---  contains a new knot invariant $\rho^{III}$  that does not appear in the BF theory. This seems to be caused by the special structure of the commutation  algebra  of the  $ISU(2)$ generators.   

\subsection{Framing dependence}

 Up to terms of order $\hbar^3$,  the normalized trace of the expectation value of the knot holonomy in the  BF theory is given by the sum $\sum_{n=0}^3 \langle W_C \rangle^{(n)}$ and can be written as    
\be
\langle W_C \rangle = e^{-i \hbar \ell k (C,C_f) [\Lambda r - (g/2) \Lambda^2] } \left [ 1 + \hbar^2 \Lambda^2 \rho_C  \,\right ] + {\cal O} (\hbar^4) \; .  
\label{8.16}
 \ee 
Expression (\ref{8.16}) is in agreement with the general structure of the BF knot invariant, in which the whole dependence of $\langle W_C \rangle$ on the framing $C_f$ of the knot $C$ is given by the overall multiplicative factor  
\be
\hbox{ BF framing factor } = e^{-i \hbar \ell k (C,C_f) [\Lambda r - (g/2) \Lambda^2] } \; . 
\label{8.17}
\ee
Let us recall that, in the Chern-Simons theory, the framing factor \cite{GMM,E} of the knot invariants is given by  
\be
\hbox{ CS framing factor } = e^{-i \frac{ \hbar }{ 2g } \ell k (C,C_f) C_2 (R) } \; , 
\label{8.18}
\ee
 where $C_2(R)$ denotes the value of the quadratic Casimir operator in the  $R$ representation ---of the structure group---   which is associated with the knot, and $g = (k / 4 \pi ) $ is the CS coupling constant \cite{ES} which multiplies the Chern-Simons action.  

The framing dependence of the knot observables has a common origin in both the BF and the CS theories.  

\begin{prop}
The {\rm BF} knot invariant $\langle W_C \rangle$ of the framed knot $C$ has the form 
\be
\langle W_C \rangle = e^{-i \hbar \ell k (C,C_f) [\Lambda r - (g/2) \Lambda^2] }  Q_C \; , 
\label{8.19}
\ee
where $Q_C$ does not depend on the framing $C_f$ of the knot $C$. 
\end{prop}

\noindent{\bf Proof.} Let us recall that the framing of the knot $C$ can be defined by means of a knot $C_f$ which belongs to the boundary of a tubular neighbourhood of $C$. If $C$ is oriented, the orientation of $C_f$ is chosen to agree with the orientation of $C$. 

It should be noted that the choice of a framing of a knot $C \subset {\mathbb R}^3$ is equivalent to the specification of a  trivialisation \cite{R} of a tubular neighbourhood $N$ of $C$. The space   $N \subset {\mathbb R}^3$ is  a solid torus, in which $C$ is the core of $N$ and $C_f \subset \der N $.   Let us define the standard solid torus $V$ as the product $V = S^1 \times D^2$, where   
the two-dimensional disc $D^2$ is represented by the unit disc in the complex plane with coordinates $\{ r e^{i \theta} \}$ in which $0 \leq r \leq 1$ and $0 < \theta \leq 2 \pi $. Let $\{ e^{i \phi }, r e^{i \theta} \}$ be coordinates of $V$; 
the standard longitude $\lambda $ of $V$ is the curve on the boundary $\der V$ of coordinates $\{ e^{i \phi} , 1 \}$ with $0 < \phi \leq 2 \pi$.  A framing for $C$ is  a homeomorphism $f : V \rightarrow N$, and the image of  $\lambda $ is precisely the knot $C_f$. 

Up to ambient isotopy, the homeomorphism $f : V \rightarrow N$ is uniquely specified  by the linking number of $C$ and $C_f$. This means that, in the quantum field theory context  of the BF or CS theories,   the whole dependence of  $\langle W_C \rangle $ on the framing is given precisely by the sum of all the perturbative contributions which are proportional to the linking number $\ell k (C, C_f)$. 

The linking number $\ell k (C, C_f)$ is given the integral  along $C$ and $C_f$ of the corresponding Gauss density which appears in the expression (\ref{2.12}) of the  components of the propagator for the connection. The propagator corresponds  to the two-point function of the connection fields  
 \be
\langle   A^a_\mu (x) B^b_\nu(y)  \rangle = \frac{i  \delta^{ab}}{4 \pi } \epsilon_{\mu \nu \lambda} \frac{(x-y)^\lambda}{|x-y|^3} \quad , \quad \langle   B^a_\mu (x) B^b_\nu(y)  \rangle = \frac{-i g \delta^{ab}}{4 \pi } \epsilon_{\mu \nu \lambda} \frac{(x-y)^\lambda}{|x-y|^3} \; , 
\label{8.20}
\ee
 that, in the BF and CS theories, receives no loop corrections (see Section~4 and \cite{GMM,ES}).    
When the components of the connection are coupled with  classical sources $J^a_\mu (x)$ and $K^a_\mu (x)$, the set of the corresponding  Feynman diagrams  is described by the generating functional  
$$  \langle   e^{  i \int d^3x (J^a_\mu A^a_\mu + K^a_\mu  B^a_\mu)  }  \rangle 
$$ 
and,  since the two-point function is connected, the sum of all the contributions containing the  linking number $\ell k (C, C_f)$ is precisey the exponential of the two-point function \cite{PS,IZ,JON}. 
This means that, by  neglecting the commutators between the  generators $J^a$ and $P^a$, the entire framing dependence of  $\langle W_C \rangle$  is given by the overall multiplicative  factor which is just the exponential of  $\ell k (C, C_f)$ multiplied by the quadratic Casimir operator which is defined by the two-point function of the connection 
\be
\hbox{framing factor } = e^{-i \hbar \ell k (C,C_f) [ PJ - (g/2) P^2] } \; . 
\label{8.21}
\ee

Let us now take into account the fact that the generators 
$\{  J^a , P^b \}$ do not generally commute.  The holonomy $h_C$ is defined by means of the path-ordered exponential and, in the perturbative expansion (\ref{7.2}) of $h_C$ in powers of the fields, the path-ordering  determines the precise position of  the $J^a$    and $P^b$  operators in the product of the group generators along the knot $C$. Let us consider the Feynman diagrams --contributing to $\langle W_C \rangle$---  in which a ${\cal A} {\cal A} $   propagator connects two points of the knot $C$. There are only two possibilities: (a) the associated group generators  are  placed in consecutive positions in the path-ordering, or (b) the associated generators are nonconsecutive.  

 \vskip 0.6 truecm

\centerline {
\begin{tikzpicture} [scale=0.8] [>=latex]
%
\draw [very thick  ] (-4,0) -- (5,0); 
\draw [very thick  ] (-4,-0.1) -- (5,-0.1); 
\draw [very thick]  (1.8,0) arc (0:180:1.3 and 1); 
\draw [very thick]  (-3,0) -- (-3,-1.5); 
\draw [very thick]  (-2,0) -- (-2,-1.5);
\draw [very thick]  (2.8,0) -- (2.8,-1.5);
\draw [very thick]  (4,0) -- (4,-1.5);
\node at (-1.08,0.32) {$\alpha$};
\node at (2.12,0.32) {$\beta$};
\end{tikzpicture}
}

\vskip 0.3 truecm
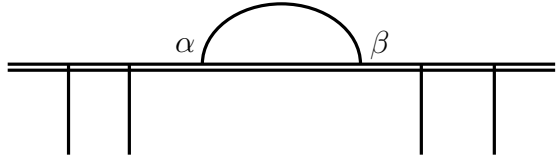
\centerline {{Figure 12.} {Part of a diagram with one  propagator associated with two consecutive generators.}}

\vskip 0.4 truecm

\noindent In the case (a), sketched in  Figure~12, the two-point function is proportional to the contraction $\WT{T^\alpha T}\null^{\! \! \beta}$ which is equal to the Casimir operators $JP$ or $P^2$, which commute with all the remaining generators and therefore  behave as  classical numbers (or classical sources).   

In case (b), depicted in Figure~13, the generators $T^\alpha$ and $T^\beta $ which are associated with the propagator are nonconsecutive, and one has, for instance, the sequence  
$T^\alpha T^\sigma T^\gamma T^\beta $; this product can be written as 
\be
T^\alpha T^\sigma T^\gamma T^\beta = T^\sigma T^\gamma \, T^\alpha T^\beta  + \left [  T^\alpha , T^\sigma T^\gamma \right ] T^\beta \; .   
\label{8.22}
\ee
The first term on the r.h.s. of expression (\ref{8.22}) contains the quadratic Casimir operator entering $\WT{T^\alpha T}\null^{\! \! \beta}$ (which is equal to $JP$ or $PP$) and, when one combines all the terms of this type with the terms coming from   case (a), one gets precisely the exponentiation  shown in equation (\ref{8.21}). 

Since the set of  all the perturbative contributions to $\langle W_C \rangle$ takes the form of a sum of knot invariants, if one extract the knot invariant $\ell k (C, C_f)$ the remaining terms necessarily represent  knot invariants. 
Thus the remaining contributions, which contain the commutator appearing in expression (\ref{8.22}), combine to produce knot invariants, which necessarily are not proportional to the linking number $\ell k (C, C_f)$ because  they do not contain the complete line integral along $C$ and $C_f$ of the Gauss density.    

 \vskip 0.6 truecm

\centerline {
\begin{tikzpicture} [scale=0.8] [>=latex]
%
\draw [very thick  ] (-4,0) -- (5,0); 
\draw [very thick  ] (-4,-0.1) -- (5,-0.1); 
\draw [very thick]  (1.8,0) arc (0:180:1.3 and 1); 
\draw [very thick]  (-2.5,0) -- (-2.5,-1.5); 
\draw [very thick]  (-0.2,0) -- (-0.2,-1.5);
\draw [very thick]  (1,0) -- (1,-1.5);
\draw [very thick]  (3.5,0) -- (3.5,-1.5);
\node at (-1.08,0.32) {$\alpha$};
\node at (2.12,0.32) {$\beta$};
\node at (-0.56,-0.45) {$\sigma$};
\node at (1.3,-0.45) {$\gamma$};
\end{tikzpicture}
}

\vskip 0.3 truecm
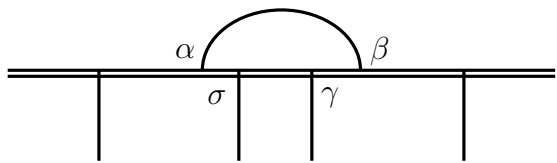
\centerline {{Figure 13.} {Part of a diagram with one  propagator associated with nonconsecutive generators.}}

\vskip 0.4 truecm

\noindent Therefore the framing dependence of $\langle W_C \rangle$ is  given by an overall  factor which is precisely the exponential of  $\ell k (C, C_f)$ multiplied by the quadratic Casimir operator which is defined by the two-point function of the connection.  In the CS theory, the quadratic Casimir operator is exactly $T^b T^b = c_2 (R)$, whereas in the BF theory the two points function gives the combination $\left [ \, PJ - (g/2) PP \, \right ] $ of Casimir operators.  {\hfill \cvd} 

\section{Conclusions}

The gauge theory of topological type which is usually called the BF theory is a superrenormalizable quantum field theory in ${\mathbb R}^3$.   We have described the structure of the Feynman diagrams which enter the perturbative expansion of the correlation functions of the connection,  the corresponding generating functional has been computed and the relationship with the Chern-Simons theory has been produced.  We have presented the complete  renormalization of the BF theory, which involves the two-points function and  three-points function of the connection.   By means of the  renormalization procedure in the space of coordinates ---which  is in complete agreement with the renormalization procedure in momentum space---   one finds that,  as in the case of the Chern-Simons theory, the two-points function of the connection does not receive loop corrections and therefore the bare propagator coincides with the dressed propagator. 

We have defined gauge invariant observables by means of appropriately normalized traces of the holonomies which are associated with oriented, framed and coloured knots in ${\mathbb R}^3$.  The colour of a knot is specified by a given  unitary  irreducible representation of the structure group $ISU(2)$. We have described   the unitary $ISU(2)$ representations with Casimir operators $P^2 = \Lambda^2$ and $JP = r \Lambda $ ---with $r=0, 1/2$--- and the $ISU(2)$ conjugacy classes have been determined.   It has been shown that the expectation value of a knot holonomy is a function of the Casimir operators of the gauge group, so  the expectation value of the normalized   trace of knot holonomies are well defined and are gauge invariant.  

The perturbative computation of the observables has been successfully achieved  up to the third order in powers of $\hbar$. The knot invariants that we have found at first and second order correspond to the knot invariants that also appear in the Chern-Simons theory. Whereas  the BF and CS knot invariants differ at the third order of perturbation theory. We have shown that the entire framing dependence of the knot observables is completely determined  by an overall multiplicative factor which is the exponential of the linking number between the knot and its framing multiplied by the combination of the quadratic Casimir operators which is determined by the two point function of the connection.  

In the present article, we have described the fundamentals of the perturbative approach to the BF theory in the case of structure group $ISU(2)$. The extensions to more complicated groups appear to be quite natural. In particular, our results admit  rather simple generalizations to the case of gauge group $ISO(2,1) $, which is related to  a gravitational model in $(2+1)$ dimensions.

\end{document}